\documentclass[aps,prd,showpacs,amsmath,amssymb,nofootinbib,twocolumn]{revtex4}
\usepackage{graphics}
\usepackage{xcolor}
\usepackage{amssymb}
\usepackage{amsmath}
\usepackage{cases}
\usepackage{epsfig,subfigure}
\usepackage{graphicx}
\usepackage{epstopdf}
\usepackage{hyperref}
\usepackage{url}

\begin{document}



\title{Thick branes with inner structure in mimetic $f(R)$ gravity}


\author{Jing Chen$^{a\,b\,c}$}
\author{Wen-Di Guo$^{a\,b\,c\,d}$}
\author{Yu-Xiao Liu$^{a\,b\,c}$
	\footnote{liuyx@lzu.edu.cn, corresponding author}}

\affiliation{$^{a}$Lanzhou Center for Theoretical Physics and Key Laboratory of Theoretical Physics of Gansu Province, Lanzhou University, Lanzhou, Gansu 730000, China\\
$^{b}$Institute of Theoretical Physics and Research Center of Gravitation, Lanzhou University, Lanzhou 730000, P. R. China\\
$^{c}$Joint Research Center for Physics, Lanzhou University and Qinghai Normal University, Lanzhou 730000 and Xining 810000, P. R. China\\
$^{d}$Centro de Astrof\'{\i}sica e Gravita\c c\~ao  - CENTRA,
Departamento de F\'{\i}sica, Instituto Superior T\'ecnico - IST,
Universidade de Lisboa - UL,
Av. Rovisco Pais 1, 1049-001 Lisboa, Portugal
}

\begin{abstract}
In this paper, we study the structure and gravitational resonances of thick branes generated by a mimetic scalar field in $f(R)$ gravity. We obtain several typical thick brane solutions for $f(R)=R+\alpha R^2$. To study their stability, we analyze the tensor perturbation of the metric. It is shown that any thick brane model with $df/dR>0$ is stable and the graviton zero mode can be localized on the brane for each solution, which indicates that the four-dimensional Newtonian gravity can be restored. The effect of the parameter $\alpha$ on the gravitational resonances is studied. As a brane splits into multi sub-branes, the effective potential of the tensor perturbation will have an abundant inner structure with multi-wells, and this will lead to new phenomena of the gravitational resonances.
\end{abstract}



\maketitle
\section{Introduction}
In last twenty years, brane world scenario has been an attractive topic and researched widely. Arkani-Hamed, Dimopoulos, and Dvali (ADD) provided an alternative solution to gauge hierarchy problem~\cite{Arkani-Hamed1998}. In ADD model, all matter fields are confined on a four-dimensional brane which is embedded in a higher-dimensional spacetime, only gravity can propagate in the bulk. After that, Randall and Sundrum proposed two different kinds of extra dimension models, RS-1 \cite{Randall1999a} and RS-2 model \cite{Randall1999}. In RS-1 model the extra dimension is compact and warped but in RS-2 model the scale of the warped extra dimension is infinite. In RS-2 model, the four-dimensional Newtonian potential can be restored even if the scale of the extra dimension is infinite. However, the thicknesses of the brane of RS models are neglected, and this is the reason why we called them thin brane models. Combining the RS-2 model with the domain wall model \cite{Rubakov1983} one can generalize the RS-2 model to a thick brane model \cite{DeWolfe2000,Csaki2000,Gremm2000}. The localization problem of matter fields on a thick brane was studied in Refs.~\cite{DeWolfe2000,Gremm2000,Kehagias2001,Melfo2006,Almeida2009,Zhao2010,Chumbes2011,Liu2011,Xie2017,Gu2017,ZhongYuan2017,ZhongYuan2017b,Zhou2018}. In thick brane world scenario, the brane is usually generated by a canonical scalar field \cite{Liu2012,Xu2015,Yu2016,Cruz2019,Wan2020,Rosa2020}. Furthermore, various kinds of thick branes generated by multi-scalar fields were also investigated widely \cite{Bazeia2004,DeSouzaDutra2008,Dutra2015,ZhongYuan2018a,Xie2020}.
For more details of thick brane worlds and extra dimensions, see the review papers~\cite{Dzhunushaliev0904.1775,Herrera-Aguilar0910.0363,Liu2018review}.

On the other hand, {physicists should answer what cause late-time universe acceleration and what is dark matter.} To solve these issues indicated by astronomical observations, physicists provided two kinds of schemes, one is assuming the existence of extra energy and matter, the other one is modifying general relativity to match observations. {Furthermore, renormalization of gravity is another motivation to modify general relativity.} Mimetic gravity is a modified theory proposed by Chamseddine and Mukhanov \cite{Chamseddine2013}. In this theory the conformal degree of freedom of the background metric is isolated and it could mimic cold dark matter \cite{Chamseddine2014,Capela2015,Mirzagholi2015,Babichev2017,Golovnev2014,Barvinsky2014,Deruelle2014,Sebastiani2017}, or dark energy \cite{Nojiri2016,Nojiri2016a,Matsumoto2016,Casalino2018}. Barvinsky found that, for positive energy density of the mimetic fluid in cosmology, mimetic gravity is free of ghost instability~\cite{Barvinsky2014}. Then another paper confirmed  that the original mimetic gravity theory suffers from ghost instability in all generality \cite{BenAchour2016}. And using the extension of mimetic gravity the inflationary solution was studied in Ref.~\cite{Mansoori2021}. On the other hand, instead of the cold dark matter this conformal degree of freedom can also mimic the scalar field which generates the thick brane. In last few years, Y. Zhong et al. investigated thick branes generated by the mimetic scalar field in Ref.~\cite{ZhongYi2018}. And it was found that one could construct some exact solutions of brane world, and some of them have inner structure. This inner structure in a thick brane can lead to new phenomena of the gravitational resonances, which was further investigated in Ref.~\cite{ZhongYi2019}. Other recent researches can be found in Refs.~\cite{Bazeia2020,Xiang2020}.

As a simple {interesting} higher-order gravity~\cite{Stelle1977}, $f(R)$ gravity has made a great success on cosmology. For example, it can explain the inflation~\cite{Nojiri2007,Cognola2008,Nojiri2008,Huang2014,Brooker2016,Sebastiani2015} and the late-time acceleration~\cite{Capozziello2005,Amarzguioui2006,Faulkner2006,Song2007,Starobinsky2007,Hu2007,Bertolami2007,Amendola2007,Bean2007}. And the ghost stability was also studied in detail (see Refs.~\cite{DeFelice2010,Sotiriou2010}). The mimetic method mentioned above was applied to $f(R)$ gravity by Odintsov~\cite{Odintsov2014}. After that, Leon and Saridakis studied the dynamical behavior of mimetic $f(R)$ gravity~\cite{Leon2015}. Odintsov and Oikonomou investigated dark energy oscillations in this gravity \cite{Odintsov2016}. The stability of de Sitter solution in this theory was  studied~\cite{Myrzakulov2015}. Recently, another work confirmed that if the usual stability conditions of the standard $f(R)$ gravity are assumed and the Lagrange multiplier $\lambda$ which related to mimetic field energy density is positive the mimetic $f(R)$ gravity is stable \cite{Ganz2019}.

Thick brane world in $f(R)$ gravity has been studied widely \cite{Afonso2007,Deruelle2008a,Borzou2009,Dzhunushaliev2010,HoffDaSilva2011,Balcerzak2011,ZhongYuan2011,Carames2013,Bazeia2013,Bazeia2014a,Bazeia2015a,Chakraborty2015a,Gu2015a,ZhongYuan2016a,Cui2018,Gu2018a,Chen2018,Hashemi2018,Dzhunushaliev2019,Wang2019,Dzhunushaliev2020,Cui2020}. The stability of $f(R)$ brane on the linear tensor perturbation was first investigated in Ref.~\cite{ZhongYuan2011} and then was further considered in Refs.~\cite{ZhongYuan2016a,Gu2015a,Cui2018,Gu2018a,Chen2018}.
In these models, the brane is generated by a background scalar field.
In this paper, we would like to study the thick brane world in mimetic $f(R)$ gravity. In this gravity theory the thick brane is generated by a mimetic scalar field coming from the conformal degree of freedom of the metric. In fact, such thick brane model was studied in mimetic gravity \cite{ZhongYi2018} and mimetic $f(T)$ gravity \cite{Guo2018}. While the thin brane scenario in mimetic $f(R)$ gravity was investigated in Ref.~\cite{Nozari2019}. Here, we focus on the thick brane model in mimetic $f(R)$ gravity. We take $f(R) = R+\alpha R^2$ as an example, to study the effect of the higher order term of the scalar curvature. We construct a series of thick branes with inner structure analytically by using the conformal degree of freedom of the mimetic $f(R)$ theory. By analyzing the linear tensor perturbation of the metric, we study the stability and gravitational resonances of the system. The result shows that, these thick branes are stable under the tensor perturbation, and the zero mode of graviton can be localized on the branes. Therefore, the four-dimensional gravity can be recovered. Apart from the graviton zero mode, we also investigate the gravitational resonances, which stay on the brane for a long time. These resonances may have experimental signals in high energy collider. The resonances on $f(R)$ brane were studied in Refs.~\cite{Xu2015,Yu2016,Zhou2018}. Compared to these works, the branes here support more gravitational resonances thanks to their inner structure. The gravitational resonances on sub-brane structure are also studied.

The organization of our work is as follows. In Sec.~\ref{sec_2}, we review mimetic $f(R)$ theory and construct three flat thick brane models with $f(R)=R+\alpha R^2$. In Sec.~\ref{sec_3}, we investigate the tensor perturbation and the graviton zero mode in each brane model. In Sec.~\ref{sec_4} we study the gravitational resonances, and show the abundant behavior of these resonances due to inner structure of the branes. Finally, in Sec.~\ref{sec_5} we come to the conclusions and discussions.

\section{Mimetic $f(R)$ gravity and thick brane models}\label{sec_2}

First, we give a brief review of thick brane in mimetic $f(R)$ theory. The original mimetic scalar comes from the conformal degree freedom of the metric \cite{Chamseddine2013}. Later, it was found that this method is equivalent to a Lagrange multiplier method that first investigated in Ref.~\cite{Lim2010}. In this paper, we will use the Lagrange multiplier method. The action of mimetic $f(R)$ gravity in five-dimensional spacetime is
\begin{equation}
	S=\int d^4xdy\sqrt{-g}
      \left[\frac{M_*^3}{2} f(R)+L(\phi)
      \right],\label{action}
\end{equation}
where $M_*$ is the fundamental mass scale, $f(R)$ is a function of the scalar curvature $R$. In this paper, we set $M_*=1$. The Lagrangian of the mimetic scalar field is
\begin{equation}
	L(\phi)=\lambda\left[g^{MN}\partial_M\phi\partial_N\phi-U(\phi)\right]-V(\phi),
\end{equation}
where $\lambda$ is the Lagrange multiplier, {$U(\phi)$ and $V(\phi)$ are two scalar potentials while the former is in fact determined by the kinetic term of the mimetic scalar field}. The field equations can be obtained by varying action \eqref{action} with respect to the metric $g_{MN}$, the scalar field $\phi$, and the Lagrange multiplier $\lambda$, respectively:
\begin{align}
	f_R R_{MN}-\frac{1}{2}g_{MN}f +\left(g_{MN}\Box^{(5)}-\nabla_M \nabla_N\right)f_R= \nonumber\\
	g_{MN}L(\phi)-2\lambda \partial_M \phi \partial_N \phi &, \label{eomgmn}\\
	2\lambda \Box^{(5)}\phi+2\nabla_M \lambda \nabla^M\phi+\lambda \frac{\partial U}{\partial \phi}+\frac{\partial V}{\partial \phi}=0&,\label{eomphi}\\
	g^{MN}\partial_M \phi \partial_N \phi - U(\phi)=0&,\label{eomlambda}
\end{align}
	where $\Box^{(5)}=g^{MN}\nabla_M \nabla_N$ is the five-dimensional d'Alembert operator and $f_R\equiv df(R)/dR$. The indices $M,N,\cdots=0,1,2,3,5$ and $\mu,\nu,\cdots=0,1,2,3$ denote the bulk and brane coordinates, respectively. In the original mimetic gravity, $U(\phi)=-1$ \cite{Chamseddine2013}. Later, it was generated as $U(\phi)\neq -1$ by Astashenok, Odintsov and Oikonomou~\cite{Astashenok2015}. In order to generate a thick brane, the mimetic scalar field should be spacelike and only depend on the extra dimension $y$ \cite{ZhongYi2018}. So from Eq. \eqref{eomlambda} we know that, $U(\phi)=g^{MN}\partial_M \phi \partial_N \phi >0$.

In this paper we consider the four-dimensional flat brane with the metric
\begin{equation}
	ds^2=a^2(y)\eta_{\mu \nu} dx^{\mu} dx^{\nu}+dy^2,
\end{equation}
	where $a(y)$ is the warp factor, $y$ represents the extra dimension, and $\eta_{\mu \nu}$ is the four-dimensional Minkowski metric. With this metric, Eqs.~\eqref{eomgmn}-\eqref{eomlambda} can be written as
\begin{align}
	f+\frac{6a'^2}{a^2}f_R+\frac{2a''}{a}f_R-\frac{6a'}{a}f'_R-2f''_R-2V& \nonumber \\
	\quad -2\lambda\left(U-\phi'^2\right)=&~0,\label{eomg1}\\
	f+\frac{8a''}{a}f_R-\frac{8a'}{a}f'_R-2V-2\lambda \left(U+\phi'^2\right)=&~0,\label{eomg2}\\
	\frac{\partial V}{\partial \phi}+2\lambda'\phi'+\lambda \left(\frac{\partial U}{\partial \phi}+8\phi'\frac{a'}{a}+2\phi''\right)=&~0,\label{eomgphi}\\
	U(\phi)-\phi'^2=&~0,\label{eomgU}
\end{align}
and
\begin{equation}
    R=-12\frac{a'^2}{a^2}-8\frac{a''}{a},\label{RicciScalar}
\end{equation}
where the symbol prime denotes the derivative with respect to the extra dimension coordinate $y$. Combining Eqs. \eqref{eomg1}, \eqref{eomg2}, and \eqref{eomgU}, we can get
\begin{align}
	&\lambda=\frac{1}{\phi'^2}\left(\frac{3a''}{2a}f_R -\frac{3a'^2}{2a^2}f_R-\frac{a'}{2a}f'_R+\frac{1}{2}f''_R \right),\label{lambda}\\
	&V=\frac{f}{2}+\frac{3a'^2}{a^2}f_R+\frac{a''}{a}f_R-\frac{3a'}{a}f'_R-f''_R.\label{eomV}
\end{align}
Note that, there are only three independent equations in Eqs.~\eqref{eomg1}-\eqref{eomgU}, because the left side of Eq. \eqref{eomgmn} is divergent-free \cite{Sotiriou2010}. Since there are six functions: $f(R)$, $a(y)$, $\phi(y)$, $\lambda(\phi)$, $V(\phi)$, and $U(\phi)$, we should fix three of them. In this paper, we will give $f(R)$, $a(y)$, and $\phi(y)$, and solve $\lambda(\phi)$, $V(\phi)$, and $U(\phi)$.

We consider $f(R)=R+\alpha R^2$ as an example, where $\alpha$ is the parameter which measures the degree of deviation from mimetic gravity. In this formula of $f(R)$ we require
\begin{equation}
	f_R > 0 \label{condition}
\end{equation}
in the whole extra dimension in order to guarantee that the graviton is not a ghost \cite{Sotiriou2010}. Then we will give three kinds of warp factors to study behavior of gravitational resonances.  There are several reasons why we choose these warp factors and mimetic scalar fields.
\begin{itemize}
	\item First, usually gravity can be localized on the brane in a five-dimensional asymptotic AdS spacetime \cite{Liu2018review}.
	
	From Eq.~\eqref{RicciScalar} we know that, at infinity, that is $y \rightarrow \pm \infty$, when $a(y) \rightarrow e^{-k|y|}$, the Ricci scalar approach to a negative constant $-20 k^2$. So if the warp factor has this kind of asymptotic behavior, the five-dimensional spacetime is an asymptotic AdS spacetime. The graviton zero mode of the tensor perturbation can be localized on the brane, which results in recovering of the four-dimensional Newtonian potential.
		
	\item Second, the localization of scalar field is another point needed to be considered. It can be shown that the scalar field can also be localized on the brane embedded in a five-dimensional asymptotic AdS spacetime.

	\item Last, thanks to the conformal degree of freedom of mimetic $f(R)$ theory, we could construct branes with inner structure analytically using different warp factors. We will construct three kinds of warp factor to get corresponding branes with inner structure next. On this basis, we can further study the new physics of brane with inner structure.
\end{itemize}

\subsection{Model 1}

\begin{figure}[!htb]
	\begin{center}
		\subfigure[The warp factor \label{model1a}]{\includegraphics[width=4.2cm]{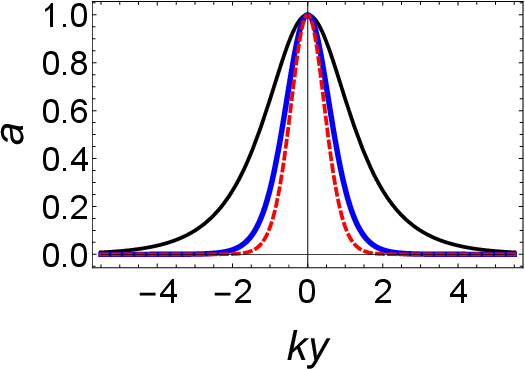}}
		\subfigure[The scalar field \label{model1b}]{\includegraphics[width=4.2cm]{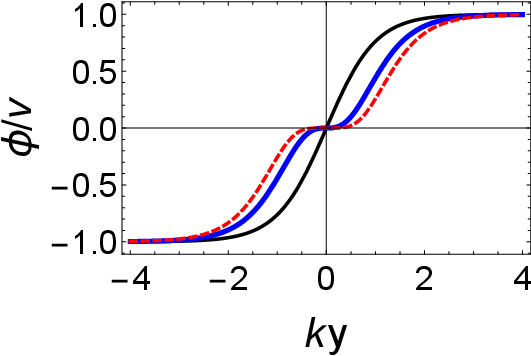}}
	\end{center}
	\caption{Shapes of the warp factor $a(y)$ and the scalar field $\phi(y)$ in model 1. The parameters are set as $n=1$ for the black lines, $n=3$ for the thick blue lines, and $n=5$ for the dashed red lines.\label{model1}}
\end{figure}

In the first model, the warp factor and the scalar field are set as follows
\begin{align}
	&a(y)=\text{sech}^{n} (ky),\\
	&\phi(y)=v\, \text{tanh}^n(ky),
\end{align}
where the parameter $k$ is usually taken as the order of $M_*$, $n$ is a positive odd integer, and $v$ is a positive parameter. Shapes of the warp factor and the scalar field for different $n$ are depicted in Figs.~\ref{model1a} and \ref{model1b}, respectively. From Fig.~\ref{model1b}, we can see that the scalar field is a single-kink and a double-kink for $n=1$ and $n\geq 3$, respectively. Besides, it can be noticed that as $n$ increases the platform near $y=0$ of the double-kink scalar field becomes wider and the warp factor becomes more concentrated.

We can solve $U$, $V$, and $\lambda$ from Eqs. \eqref{eomgU}-\eqref{eomV} as a function of $\phi$,
\begin{align}
		U(\phi)=&~k^2 n^2 v^2 \tilde{\phi} ^{\frac{2(n-1)}{n}}\left(  \tilde{\phi}^{\frac{2}{n}}-1 \right)^2,\\
		V(\phi)=&~k^2 n\biggl(8 \alpha  k^2 (n+1) (n+6) (5 n+2)\tilde{\phi}^4\nonumber \\
		&~ - \left(8 \alpha  k^2 (n (37 n+56)+16)+6 n+3\right)\tilde{\phi}^2\nonumber \\
		&~ + 32 k^2 (\alpha +3 \alpha  n)+3\biggr),\\
  \lambda(\phi)=&~ \frac{\tilde{\phi}^{2-2 n}}
                        {2 n v^2 \left(1-\tilde{\phi}^2 \right)^2}
                   \biggl(8 \alpha  k^2 (5 n+2) (n (3 n+2)-6) \tilde{\phi}^4\nonumber \\
		&~ + \left(3-32 \alpha  k^2 (n (4 n-9)-4)\right)\tilde{\phi} ^2 \nonumber \\
		&~ -\left(16 \alpha  k^2 (5 n+2)+3\right)\biggr),
\end{align}
where $\tilde{\phi}=\frac{\phi}{v}$.

\begin{figure}[h]
	\begin{center}
		\subfigure[$n=1$\label{density1_1}]{\includegraphics[width=4.2cm]{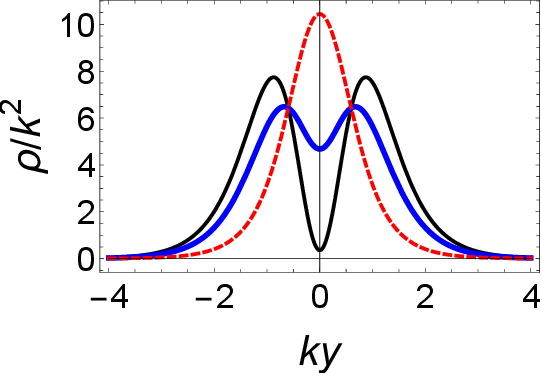}}
		\subfigure[$n=3$\label{density1_2}]{\includegraphics[width=4.2cm]{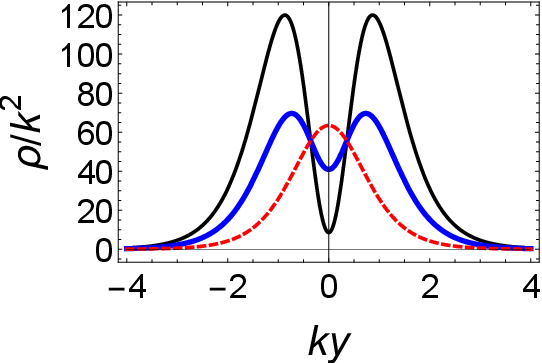}}
	\end{center}
	\caption{The energy density of model 1 with $v=1$, and $n=1$ (the left panel) and $n=3$ (the right panel). In the left panel we set $k^2 \alpha=-0.06$ for the black line, $k^2 \alpha=-0.03$ for the thick blue line, and $k^2 \alpha=0.01$ for the dashed red line. In the right panel we set $k^2 \alpha=-0.018$ for the black line, $k^2 \alpha=-0.0073$ for the thick blue line, and $k^2 \alpha=0.00014$ for the dashed red line.\label{density1}}
\end{figure}

Considering the condition \eqref{condition} in this model, we obtain the range of $k^2 \alpha$:
\begin{align}
	-\frac{1}{16n}<k^2 \alpha<\frac{1}{40n^2}. \label{range1}
\end{align}
Here we show the shapes of the dimensionless energy density $\rho /k^2$ in Fig.~\ref{density1} for several situations. Besides, it is found that the brane will spilt into two sub-branes when $-\frac{1}{16n}<k^2 \alpha <-\frac{1}{8 n(4+5n)}$.

\subsection{Model 2}

\begin{figure}
	\begin{center}
		\includegraphics[width=6.2cm]{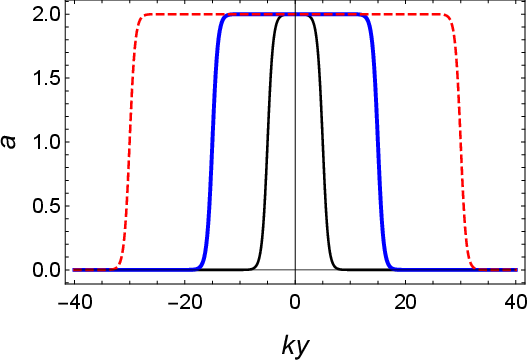}
		\caption{The warp factor of model 2. The parameter $v$ is set to $v=1$, and $kb=5$ for the black line, $kb=15$ for the thick blue line, and $kb=30$ for the dashed red line.\label{model2}}
	\end{center}
\end{figure}

\begin{figure}
	\begin{center}
		\includegraphics[width=6.2cm]{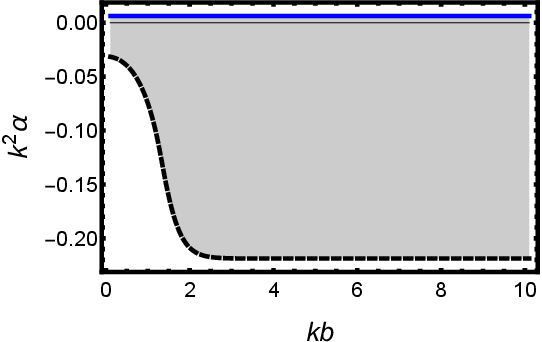}
		\caption{Plot of the range (the dark region) of $k^2 \alpha$ and $kb$ in the condition of $f_R>0$ for model 2.\label{alpha}}
	\end{center}
\end{figure}

\begin{figure}
	\begin{center}
		\subfigure[$kb=7$ \label{density2_1}]{\includegraphics[width=4.2cm]{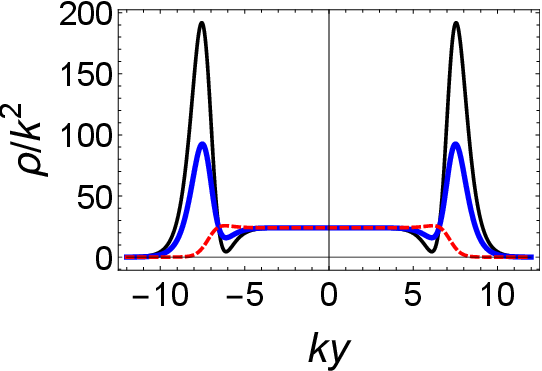}}
		\subfigure[$k^2 \alpha=-0.008$ \label{density2_2}]{\includegraphics[width=4.2cm]{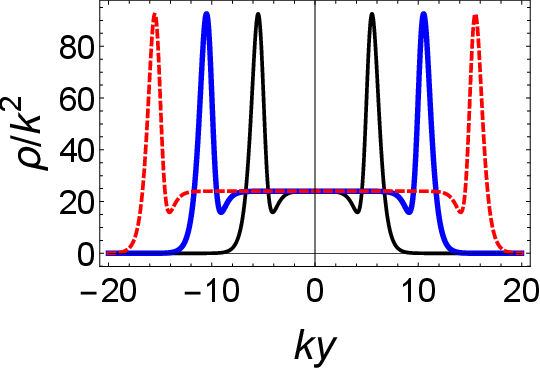}}
	\end{center}
	\caption{The energy density of model 2. The parameter $v$ is set to $v=1$. In the left panel, we set $kb=7$, and $k^2 \alpha =-0.018$ for the black line, $k^2 \alpha=-0.008$ for the thick blue line, and $k^2 \alpha=0.006$ for the dashed red line. In the right panel, we set $k^2 \alpha=-0.008$, and $kb=5$ for the black line, $kb=10$ for the thick blue line, and $kb=15$ for the dashed red line.\label{density2}}
\end{figure}

In the second model, the warp factor has a platform near the origin of the extra dimension, which can be seen from Fig.~\ref{model2}. The scalar field is a single-kink configuration. The expressions of the warp factor, the scalar field, and the scalar potential $U$ are given by
\begin{align}
	 a(y)&=\text{tanh} [k(y+b)]-\text{tanh} [k(y-b)], \\
	 \phi(y)&=v\, \text{tanh}(ky),\\
	 U(\phi)&=\frac{k^{2}}{v^{2}}\left(\phi^{2}-v^{2}\right)^{2}.
\end{align}
Here we do not show the complicated $V(\phi)$ and $\lambda(\phi)$, which can be solved analytically. Note that the width of the platform of the warp factor is controlled by the parameter $kb$. Considering the condition \eqref{condition}, we can obtain the range of the parameter $k^2 \alpha$ numerically, which is shown in Fig.~\ref{alpha}. From this figure, we can see that when $kb\gtrsim 2$ the range of $k^2 \alpha$ is of about
\begin{equation}
	-0.219<k^2 \alpha<0.00625.\label{rangec1}
\end{equation}
 The dimensionless energy density $\rho /k^2$ in this model is shown in Fig.~\ref{density2}. The brane will split into multi-branes for $kb\gtrsim 2$ with $k^2 \alpha$ in the range of \eqref{rangec1}. When $kb$ close to zero the warp factor and the energy density will close to model 1 with $n=1$. In other words, whether the brane split or not depends on the value of $k^2 \alpha$, when $k^2 \alpha$ is smaller than some value the brane will split into two sub-branes.

We can also extend this warp factor to the form as follows
\begin{align}
	a(y)=&~\text{tanh}[k(y-b-d)]-\text{tanh}[k(y+b+d)]\nonumber \\
	&-\text{tanh}[k(y+d)]+\text{tanh}[k(y-d)],
\end{align}
where $b$ and $d$ are positive parameters. Using this warp factor one can construct a brane world with double number of sub-branes, which will be further investigated in Sec. \ref{sec_4}.
	
\subsection{Model 3}
\begin{figure}[h]
	\begin{center}
		\subfigure[The warp factor \label{model3}]{\includegraphics[width=4.2cm]{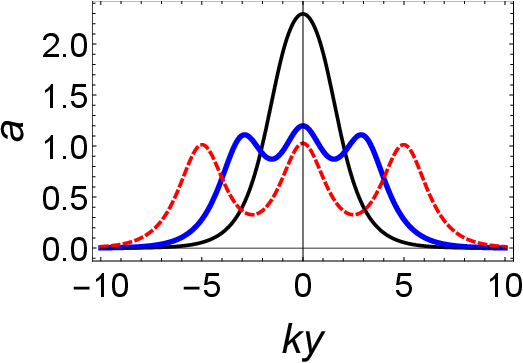}}
		\subfigure[The { energy density} \label{density3}]{\includegraphics[width=4.2cm]{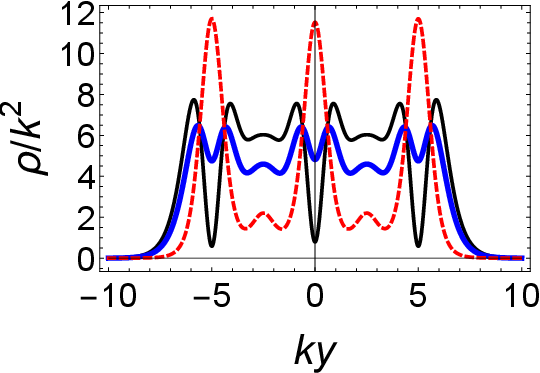}}
	\end{center}
	\caption{The warp factor and the energy density of model 3. The parameter $v$ is set to $v=1$. In the left panel we set $kb=1$ for the black line, $kb=3$ for the thick blue line, and $kb=5$ for the dashed red line. In the right panel we set $kb=5$ and $k^2 \alpha =-0.06$ for the black line, $k^2 \alpha=-0.03$ for the thick blue line, and $k^2 \alpha=0.02$ for the dashed red line.}
\end{figure}

	The third model is
\begin{align}
	a(y)=&~\text{sech}(k(y-b))+\text{sech}(ky)\nonumber \\
	&~+\text{sech}(k(y+b)),\\
	\phi(y)=&~v\, \text{tanh}(ky), \\
	 U(\phi)=&~\frac{k^2}{v^2}(\phi^2-v^2)^2.
\end{align}
The scalar field is also a single-kink and the warp factor has peaks for large parameter $kb$, which can be seen from Fig.~\ref{model3}. Here we also do not show the complicated expressions of $\lambda(\phi)$ and $V(\phi)$.

\begin{figure}[h]
	\begin{center}
		\includegraphics[width=6.2cm]{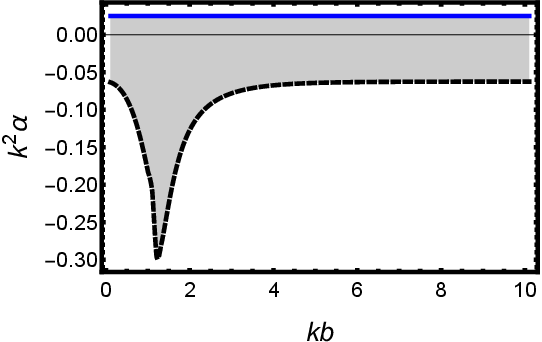}
		\caption{Plot of the range (the dark region) of $k^2 \alpha$ and $kb$ in the condition of $f_R>0$ for model 3. \label{alpha2}}
	\end{center}
\end{figure}

The range of $k^2 \alpha$ due to the condition $f_R>0$ is shown in Fig.~\ref{alpha2}. We can see that for $kb\gtrsim 4$, the range of $k^2 \alpha$ is $-0.0625<k^2 \alpha<0.025$. The dimensionless energy density $\rho /k^2$ for $kb=5$ in this model is shown in Fig.~\ref{density3}. In this situation, the sub-branes will always exist for $kb\gtrsim 4$ and $k^2 \alpha$ in the range we obtained. When $kb$ close to zero, the warp factor and the energy density will also close to model 1 with $n=1$. In other words, whether the brane split or not for the case of small $kb$ depends on the value of $k^2 \alpha$. When $k^2 \alpha$ less than some critical value the brane will split into two sub-branes.

We can also extend this three-peak model to multiple-peak model with the following warp factor
\begin{equation}
	a(y)=\sum^{N}_{n=-N} \text{sech}(k(y+nb)),
\end{equation}
where $N$ is an arbitrary positive integer.

\section{Tensor perturbations and localization}\label{sec_3}

In this section, we investigate the stability of the system by studying the tensor perturbation. The perturbed metric is given by
\begin{equation}
	g_{MN}=\bar{g}_{MN}+\delta g_{MN},
\end{equation}
with
\begin{equation}
	\bar{g}_{MN}=
	\begin{pmatrix}
		a^2(y) \eta_{\mu \nu} & 0 \\
		0 & 1
	\end{pmatrix},
	\quad \delta g_{MN}=
	\begin{pmatrix}
		a^2(y) h_{\mu \nu} & 0 \\
		0 & 0
	\end{pmatrix},
\end{equation}
where $h_{\mu \nu}=h_{\mu \nu}(x^{\rho},y)$ depends on all coordinates. Here, we only consider the tensor perturbation, so $\delta g_{5N}=0$. Using the relation $g^{NP}g_{PM}=\delta ^N_M$, one can obtain the inverse of $\delta g_{MN}$,
\begin{equation}
	\delta g^{MN}=
	\begin{pmatrix}
		-a^{-2}(y)h^{\mu \nu} & 0 \\
		0 & 0
	\end{pmatrix},
\end{equation}
where $h^{\mu \nu}=\eta^{\mu \lambda}\eta^{\nu \rho}h_{\lambda \rho}$. Note that, we only keep the first order. Then the following relations can be obtained
\begin{align}
	\delta R_{\mu \nu}=&-\frac{1}{2}\left( \square^{(4)}h_{\mu \nu}+\partial_{\mu}\partial_{\nu}h -\partial_{\nu}\partial_{\sigma}h^{\sigma}_{\mu}-\partial_{\mu}\partial_{\sigma}h^{\sigma}_{v} \right) \nonumber \\
	&-2a a'h'_{\mu
	\nu} -3h_{\mu \nu}a'^2-a h_{\mu \nu}a''-\frac{1}{2}a^2 h''_{\mu \nu} \nonumber \\
	&-\frac{1}{2}a\eta_{\mu \nu}a'h',\label{Rmunu}\\
	\delta R_{\mu 5}=&\frac{1}{2}\partial_y \left( \partial_{\lambda} h^{\lambda}_{\mu}-\partial_{\mu}h \right),\label{Rmu5}\\
	\delta R_{55}=&-\frac{1}{2}\left( \frac{2a' h'}{a}+h'' \right),\label{R55}\\
	\delta R=&-\frac{\square^{(4)}h}{a^2}+\frac{\partial_{\mu}\partial_{\nu}h^{\mu \nu}}{a^2}-\frac{a'}{a}5h'-h''.\label{R}
\end{align}
Here $\square^{(4)}=\eta^{\mu \nu}\partial_{\mu}\partial_{\nu}$ is the four-dimensional d'Alembert operator, and $h=\eta^{\mu \nu}h_{\mu \nu}$ is the trace of the tensor perturbation. Hereafter, we use the transverse-traceless gauge $h=0=\partial_{\mu}h^{\mu}_{\nu}$. Then, Eqs.~\eqref{Rmunu}-\eqref{R} can be simplified further. Using the above relations, we obtain the tensor perturbation equation
\begin{align}
	&-\frac{1}{4}a^{2}h_{\mu \nu}f(R)+\frac{1}{2}a^2 h_{\mu \nu}V-\frac{1}{2}\lambda a^2 h_{\mu \nu}\left[ \phi'^2-U(\phi) \right]\nonumber \\
	&\quad +\frac{1}{2}\left( -\frac{1}{2}\square^{(4)}h_{\mu \nu}-3a'^2 h_{\mu \nu}-2aa'h'_{\mu \nu}-aa''h_{\mu \nu}\right.\nonumber \\
	&\left. -\frac{a^2}{2}h''_{\mu \nu} \right)f_R +a^2\left[ \frac{1}{2}h_{\mu \nu}\left( 3\frac{a'}{a}f'_R+f''_R \right)-\frac{1}{4}h'_{\mu \nu}f'_R \right]=0.
\end{align}
Comparing with the Einstein equation \eqref{eomgmn}, one can easily obtain
\begin{equation}\label{perturbation}
	\left( a^{-2}\square^{(4)}h_{\mu \nu}+4\frac{a'}{a}h'_{\mu \nu}+h''_{\mu \nu} \right)f_R+h'_{\mu \nu}f'_R=0.
\end{equation}
After the coordinate transformation $dz=a^{-1}dy$, we can rewrite the perturbed equation \eqref{perturbation} as
\begin{equation}
	\left[ \partial^2_z+\left( 3\frac{\partial_z a}{a}+\frac{\partial_{z}f_R}{f_R} \right)\partial_z+\square^{(4)} \right]h_{\mu \nu}=0.
\end{equation}
Considering the Kaluza-Klein (KK) decomposition $h_{\mu \nu}(x^{\rho},z)=\left( a^{-3/2}f_R^{-1/2} \right)\epsilon_{\mu \nu}(x^{\rho})\psi(z)$, where $\epsilon_{\mu \nu}(x^{\rho})$ satisfies the transverse and traceless conditions $\eta^{\mu \nu}\epsilon_{\mu \nu}=0$ and $\partial_{\mu}\epsilon^{\mu}_{\nu}=0$, we obtain the following Schr\"odinger-like equation for the extra-dimensional part  $\psi(z)$
\begin{equation}
    \left[- \partial^2_z  + W(z) \right]\psi(z)=m^2 \psi(z),\label{Schrodingerlike}
\end{equation}
where
\begin{align}
    W(z) =&~ \frac{3}{4}\frac{(\partial _z a)^2}{a^2}+\frac{3}{2}\frac{\partial _z a ~\partial _z f_R}{a f_R}\nonumber \\
    &-\frac{1}{4}\frac{(\partial _z f_R)^2}{f^2_R}+\frac{3}{2}\frac{\partial _z^2 a}{a}+\frac{1}{2}\frac{\partial _z^2 f_R}{f_R},\label{potential}
\end{align}
is the effective potential of gravitons. The effective potential \eqref{potential} in $y$ coordinate is
\begin{equation}
   W(z(y))=\frac{9}{4}a'^2+2\frac{aa'f'_R}{f_R}-\frac{1}{4}a^2\frac{f'^2_R}{f^2_R}+\frac{3}{2}aa''+\frac{1}{2}\frac{a^2f''_R}{f_R}.\label{potentialy}
\end{equation}
It would be clearer if Eq. \eqref{Schrodingerlike} is written in the following form
\begin{equation}
	QQ^{\dagger} \psi(z)=m^2 \psi(z),\label{simplyS}
\end{equation}
	with
\begin{align}
	&Q=\partial_z+\left( \frac{3}{2}\frac{\partial_z a}{a}+\frac{1}{2}\frac{\partial_z f_R}{f_R} \right), \\
	&Q^{\dagger}=-\partial_z + \left( \frac{3}{2}\frac{\partial_z a}{a}+\frac{1}{2}\frac{\partial_z f_R}{f_R} \right).
\end{align}
Equation \eqref{simplyS} guarantees that there is no tachyonic KK graviton with $m^2<0$. So, the system is stable under the tenser perturbation.

The solution of the graviton zero mode is
\begin{equation}
	\psi^{(0)}(z)=\sqrt{N_0 a^{3}(z)f_R (z)} , \label{gravitonZeroMode}
\end{equation}
where $N_0$ is the normalization constant. { The zero mode $\psi^{(0)}(z)$ is in fact the massless graviton. It should be normalizable in order to recover the four-dimensional Newtonian gravity. Thus, we should have the following normalization condition
\begin{equation}
	\mathcal{I}\equiv\int^{\infty}_{-\infty}\left| \psi^{(0)}(z) \right|^2 dz =1.
\end{equation}
Considering the solution (\ref{gravitonZeroMode}) and the coordinate transformation $dz=a^{-1}dy$, the above condition reads as
\begin{align}
	\mathcal{I} =&\int^{\infty}_{-\infty}N_0 a^{2}(y)f_R (y) dy
	\nonumber \\
	=&\int^{\infty}_{-\infty}N_0
        \left[a^2(y)-24 \alpha a'(y)^2 -16 \alpha a(y) a''(y) \right]
        dy =1.
        \label{integrate}
\end{align}

For model 1, the graviton zero mode can be normalized and the normalization constant is given by
\begin{align}
	N_0=\frac{n(n+1)}{(8n^2 k^2 \alpha -2n -1){}_{2}F_1(1,-1-n;1+n;-1)},
\end{align}
where ${}_2F_1$ is the hypergeometric function. It can be seen that this value is positive and finite when $n$ is a positive integer.

For models 2 and 3, we have respectively
\begin{align}
	N_0^{-1}=&~
         8kb \coth(2kb)-4
        -\frac{16}{3}k^2 \alpha \text{csch}^3 (2kb)
         \Big[9\sinh(2kb)	\nonumber \\
	&        +\sinh(6kb)-24 kb \cosh(2kb)
         \Big],\\
	N_0^{-1}
	=&~ 6+4kb\text{csch}(kb)(2+\text{sech}(kb))-4k^2 \alpha \text{csch}^3(2kb)\times
	\nonumber \\
	&\Big[-24kb -8kb \Big( 22\cosh(kb)+9\cosh(3kb)
	\nonumber \\
	&+\cosh(4kb)+\cosh(5kb)\Big)+\sinh(6kb)
	\nonumber \\
	&+16\sinh(5kb)+8\sinh(4kb)+48 \sinh(3kb)
	\nonumber \\
	&-3\sinh(2kb)+32\sinh(kb)\Big].
\end{align}
}

\begin{figure}[!htb]
	\begin{center}
		\subfigure[Model 1\label{zero1}]{\includegraphics[width=4.2cm]{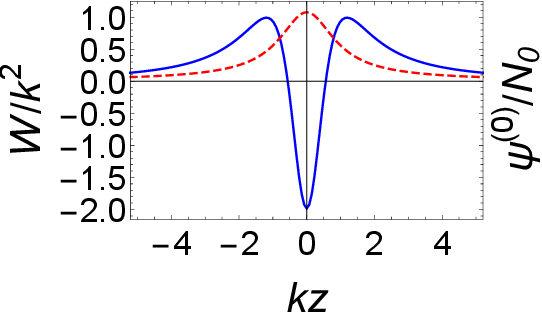}}
		\subfigure[Model 2\label{zero2}]{\includegraphics[width=4.2cm]{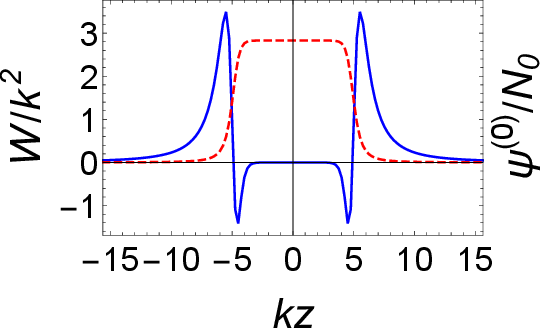}}
		\subfigure[Model 3\label{zero3}]{\includegraphics[width=4.2cm]{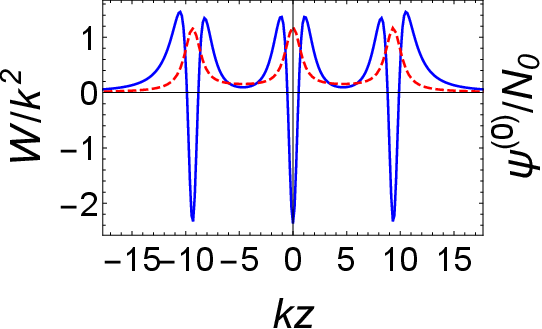}}
	\end{center}
	\caption{The effective potentials (the blue lines) and the corresponding graviton zero modes (the red dashed lines) for the three models. For model 1, we set $v=1$ and $k^2 \alpha=0.01$; for model 2, $v=1$, $kb=10$, and $k^2 \alpha=0.006$; and for model 3, $v=1$, $kb=5$, and $k^2 \alpha=0.02$.\label{zeromode}}
\end{figure}

This means that the graviton zero mode for each model considered in this paper can be localized near the brane \footnote{
Actually, the equation $QQ^{\dagger} \psi(z)=0$ has another independent solution:
	\[
	\psi^{(0)}(z)=M_0 a^{3/2}(z)f^{1/2}_R (z)\int \frac{1}{a^3(z)f_R(z)}dz,
	\]
   which was first investigated by Z.-Q. Cui, et al. in Ref.~\cite{Cui2020}.
   But this solution is divergent at $z\rightarrow \pm \infty$ for each model we considered, so it is abandoned.
	}
\cite{Csaki2000,Soda2002}. The effective potentials and the graviton zero modes for the three models are shown in Fig. \ref{zeromode}. From the analysis above of this section we notice that the tenser perturbation in mimetic $f(R)$ gravity is the same as that in $f(R)$ gravity \cite{ZhongYuan2011}. The reason is that the tenser perturbation is independent of the mimetic scalar field. And this result is also consistent with the mimetic brane world \cite{ZhongYi2018}. But the mimetic scalar field gives a degree of freedom which could generate more abundant inner structure.

\section{The gravitational resonances of mimetic $f(R)$ brane}\label{sec_4}

The abundant inner structure of the effective potential may lead to some interesting KK resonant behavior which will be studied in this section.

To get numerical solutions of Eq. \eqref{Schrodingerlike}, a convenient way is to impose the following boundary conditions for the KK modes
\begin{align}
	\psi_{\text{even}}(0)=1,&\quad \partial_z \psi_{\text{even}}(0)=0;\label{evencondition}\\
	\psi_{\text{odd}}(0)=0,&\quad \partial_z \psi_{\text{odd}}(0)=1,\label{oddcondition}
\end{align}
where $\psi_{\text{even}}$ and $\psi_{\text{odd}}$ denote even and odd parity KK modes of $\psi(z)$, respectively. The integration of the function $\left| \psi (z) \right|^2$ can be considered as the probability of finding massive KK gravitons along the extra dimension. In order to find gravitational KK resonances, one can define the relative probability \cite{Liu2009}:
\begin{equation}
	P(m^2)=\frac{\int^{z_b}_{-z_b}|\psi(z)|^2 dz}{\int^{z_{\text{max}}}_{-z_{\text{max}}}|\psi(z)|^2 dz},\label{relprobability}
\end{equation}
where $2z_b$ is the approximate width of the brane and $z_{\text{max}}$ can be taken as $10z_b$. For a given $m^2$, using conditions \eqref{evencondition} and \eqref{oddcondition}, the even parity KK mode $\psi_\text{even}$ and the odd parity KK mode $\psi_\text{odd}$ can be solved numerically from the Schr\"odinger-like equation \eqref{Schrodingerlike}. In this case, the relative probability $P$ corresponding to $\psi_{\text{even}}$ or $\psi_{\text{odd}}$ can be obtained. As a function of $m^2$, the relative probability will have a peak for some value of $m^2$. Then we may regard this mode as a gravitational KK resonance. Actually, we treat it as a KK resonance only when the corresponding peak has a full width at half maximum $\Gamma$, i.e., the width of the half height of the peak value. We can define the lifetime of the gravitational KK resonance as $\tau = \Gamma^{-1}$.

Next, we will investigate gravitational KK resonances in three cases. Before that, in order to investigate conveniently we introduce some dimensionless parameters. Using the parameter $k$, we can define the following dimensionless parameters $\bar{y}=ky$, $\bar{z}=kz$, $\bar{b}=kb$, $\bar{\tau}=k\tau$, $\bar{\Gamma}=\Gamma/k$, $\bar{m}=m/k$, {$\bar{W}=W/k^2$}, and $\bar{\alpha}=k^2 \alpha$. In our system, both parameters $\bar{b}$ and $\bar{\alpha}$ can affect the KK resonance behavior. The effects of the parameter $\bar{b}$ has been investigated in Refs.~\cite{ZhongYi2019,Fu2014,Tan2020}, so in this paper, we will focus on the effects of the parameter $\bar{\alpha}$, which can measure the deviation from mimetic gravity.

\subsection{Case 1}

\begin{figure}
	\begin{center}
		\includegraphics[width=6.2cm]{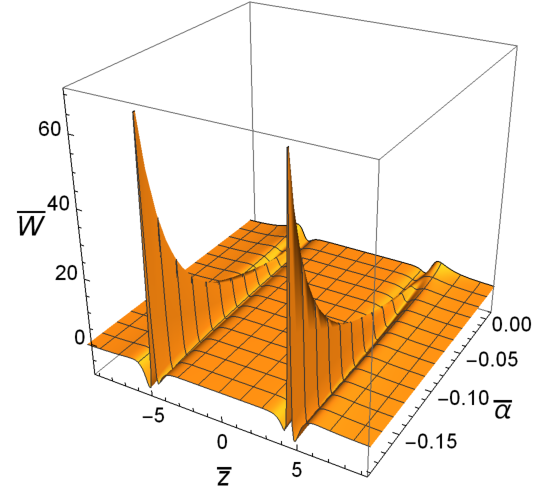}
	\end{center}
	\caption{The effect of the parameter $\bar{\alpha}$ on the effective potential with $\bar{b}=10$ for case 1.\label{poalpha1}}
\end{figure}

We begin with the following warp factor
\begin{equation}
	a(\bar{y})=\tanh(\bar{y}+\bar{b})-\tanh(\bar{y}-\bar{b}).
\end{equation}
In this case we choose $\bar{b}=10$ as a specific example to investigate gravitational KK resonances. So from Eq. \eqref{rangec1} we know that $\bar{\alpha}$ has two bounds. Effects of the parameter $\bar{\alpha}$ on the effective potential can be seen from Fig.~\ref{poalpha1}. From this figure we can see that, the height of the two big barriers decreases rapidly with $\bar{\alpha}$. When $\bar{\alpha}$ approaches to zero, the barriers will disappear. Besides, the depth of the effective potential will decrease first and then increase with $\bar{\alpha}$. {It is worth to note that in Ref.~\cite{Xu2015} the KK resonances were found when $\bar{\alpha}$ is out of the range of Eq. \eqref{condition}.}

\begin{figure}
	\begin{center}
		\subfigure[The effective potential\label{potential1_1}]{\includegraphics[width=4.2cm]{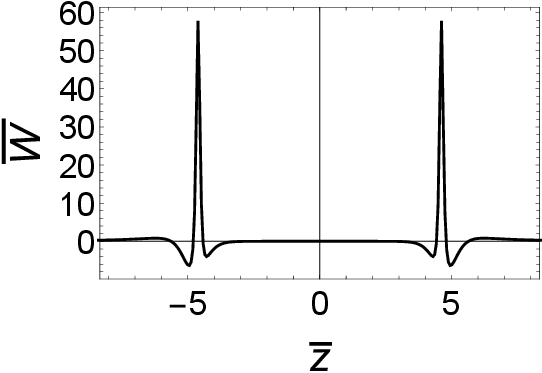}}
		\subfigure[The relative probability\label{probability1_1}]{\includegraphics[width=4.2cm]{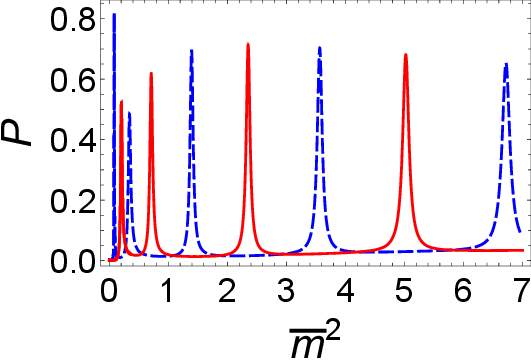}}
		\subfigure[The first odd wave function with $\bar{m}^2=0.0831$ \label{resonant1_1odd}]{\includegraphics[width=4.2cm]{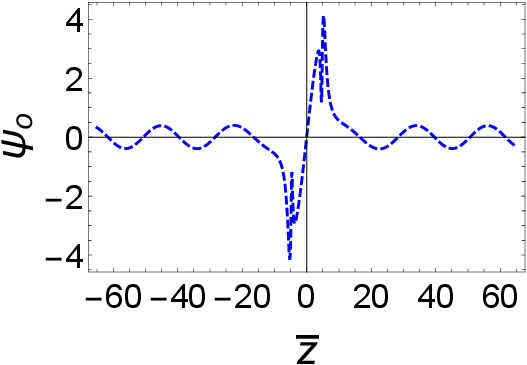}}
		\subfigure[The first even wave function with $\bar{m}^2=0.2045$ \label{resonant1_1even}]{\includegraphics[width=4.2cm]{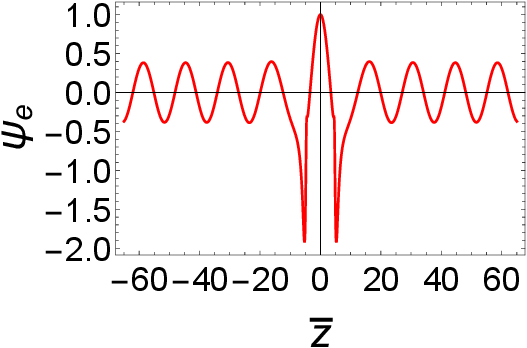}}
	\end{center}
	\caption{The effective potential, the relative probability, and the wave functions of the first odd and even KK resonances with $\bar{b}=10$ and $\bar{\alpha}=-0.19$ for case 1. \label{case1_1}}
\end{figure}

\begin{figure}
	\begin{center}
		\subfigure[The effective potential\label{potential1_2}]{\includegraphics[width=4.2cm]{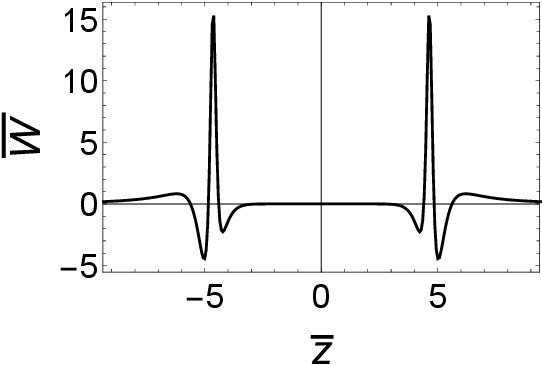}}
		\subfigure[The relative probability\label{probability1_2}]{\includegraphics[width=4.2cm]{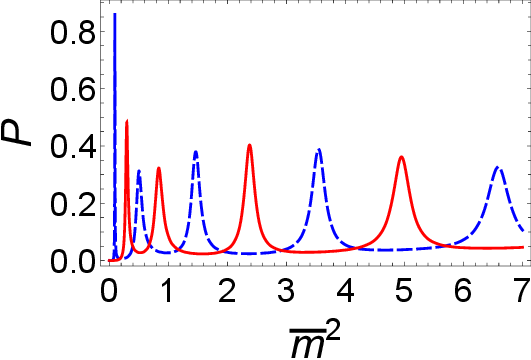}}
		\subfigure[The first odd wave function with $\bar{m}^2=0.0924$ \label{resonant1_2odd}]{\includegraphics[width=4.2cm]{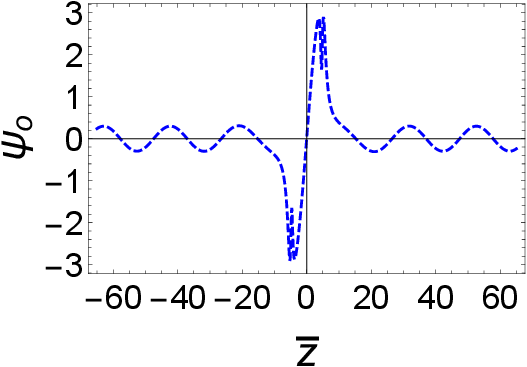}}
		\subfigure[The first even wave function $\bar{m}^2=0.2945$ \label{resonant1_2even}]{\includegraphics[width=4.2cm]{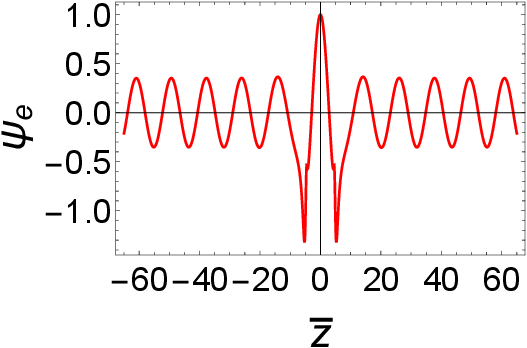}}
	\end{center}
	\caption{The effective potential, the relative probability, and the wave functions of the first odd and even KK resonances with $\bar{b}=10$ and $\bar{\alpha}=-0.14$ for case 1. \label{case1_2}}
\end{figure}

\begin{figure}
	\begin{center}
		\subfigure[The effective potential\label{potential1_3}]{\includegraphics[width=4.2cm]{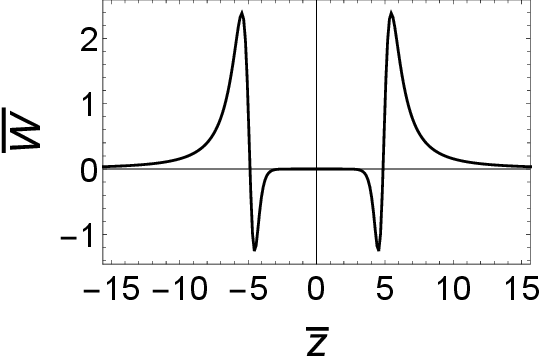}}
		\subfigure[The relative probability\label{probability1_3}]{\includegraphics[width=4.2cm]{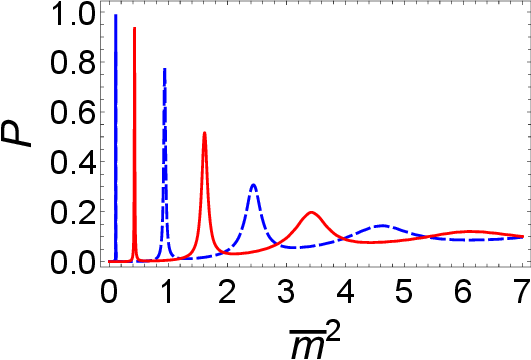}}
		\subfigure[The first odd wave function with $\bar{m}^2=0.1087$ \label{resonant1_3odd}]{\includegraphics[width=4.2cm]{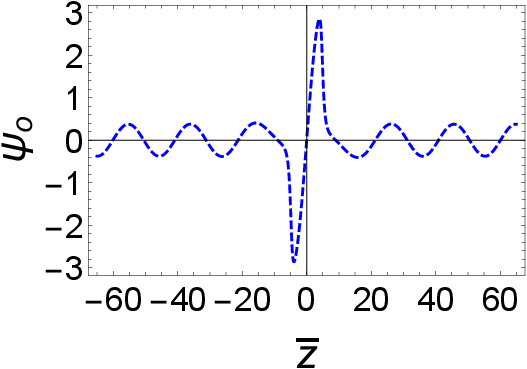}}
		\subfigure[The first even wave function with $\bar{m}^2=0.4277$ \label{resonant1_3even}]{\includegraphics[width=4.2cm]{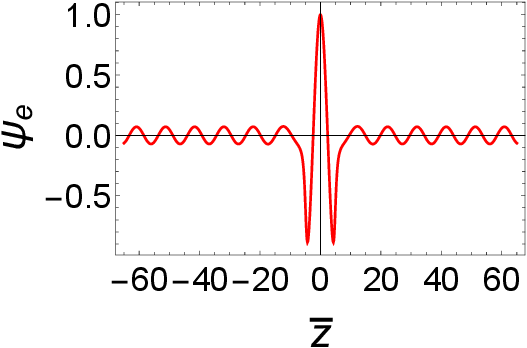}}
	\end{center}
	\caption{The effective potential, the relative probability, and the wave functions of the first odd and even KK resonances with $\bar{b}=10$ and $\bar{\alpha}=-0.0012$ for case 1. \label{case1_3}}
\end{figure}

In this case we set $\bar{z}_b$ in the relative probability \eqref{relprobability} as $\bar{z}_b=6.5$. For $\bar{\alpha}=-0.19$, $-0.14$, and $-0.0012$ we can obtain the relative probabilities. Then, the corresponding effective potentials, the relative probabilities, and the wave functions of the first odd and even KK resonances are shown in Figs.~\ref{case1_1}, \ref{case1_2}, and \ref{case1_3}. From Figs.~\ref{potential1_1}, \ref{potential1_2}, and \ref{potential1_3}, we can see that, the height of the effective potential barrier decreases with the parameter $\bar{\alpha}$. Besides, from Figs.~\ref{probability1_1} and \ref{probability1_2} we can see that the peak values do not decrease with $\bar{m}^2$ monotonously, which is an unusual phenomenon.

\begin{figure}
	\begin{center}
		\subfigure[$\bar{m}^2$\label{malpha}]{\includegraphics[width=4.2cm]{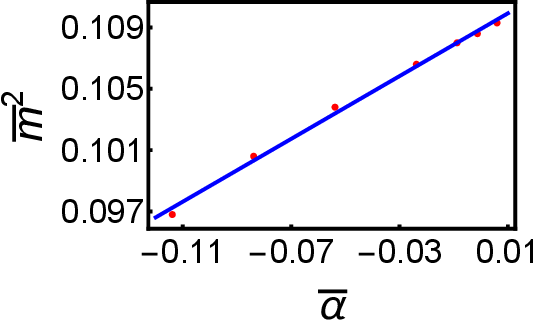}}
		\subfigure[$\bar{\tau}_1$\label{taualpha}]{\includegraphics[width=4.2cm]{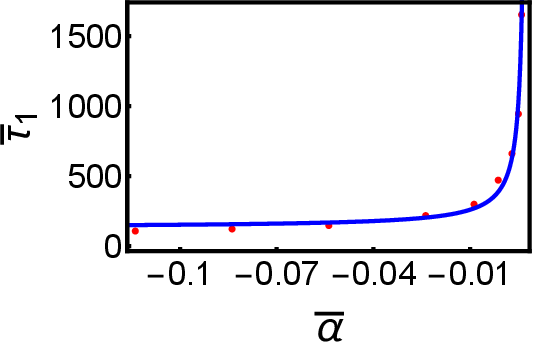}}
	\end{center}
	\caption{The effects of the parameter $\bar{\alpha}$ on the mass square $\bar{m}^2$ and the lifetime $\bar{\tau}$ of the first KK resonance for case 1. Red dots are numerical results, blue solid lines are fit functions. In the left panel the fit function is $\bar{m}^2=0.102 \bar{\alpha}+0.109$, in the right panel the fit function is $\bar{\tau}_1=\frac{1}{0.00335 -0.447 \bar{\alpha}}+131.621$.\label{fit1}}
\end{figure}

We are also interested in the first gravitational KK resonance, its lifetime $\bar{\tau}_1$ and the mass square $\bar{m}^2$ for different values of $\bar{\alpha}$ are shown in Fig.~\ref{fit1}. They can be fitted as the following two functions
\begin{align}
	\bar{m}^2&=0.102 \bar{\alpha}+0.109,\label{fitfunction1_1}\\
	\bar{\tau}_1&=\frac{1}{0.00335 -0.447 \bar{\alpha}}+132.\label{fitfunction1_2}
\end{align}
From Eq. \eqref{fitfunction1_1} we can see that the mass square $\bar{m}^2$ of the first KK resonance linearly increases with the parameter $\bar{\alpha}$. And from Eq. \eqref{fitfunction1_2} we can see that the lifetime of the first KK resonance increases with $\bar{\alpha}$ slowly first and then rapidly when $\bar{\alpha}$ closes to the upper bound.

\subsection{Case 2}

\begin{figure}
	\begin{center}
		\includegraphics[width=6.2cm]{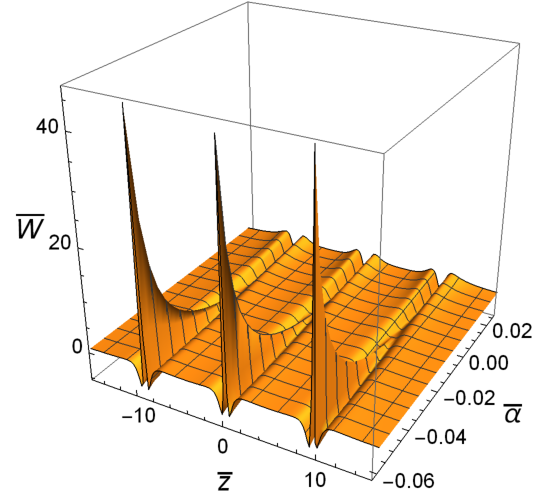}
	\end{center}
	\caption{The effect of the parameter $\bar{\alpha}$ on the effective potential with $\bar{b}=5$  for case 2. \label{poalpha2}}
\end{figure}

In this case, we investigate the brane world with three sub-branes, for which the warp factor is
\begin{equation}
	a(\bar{y})=\text{sech}(\bar{y}-\bar{b})+\text{sech}(\bar{y})+\text{sech}(\bar{y}+\bar{b}).\label{warpf2}
\end{equation}
Without loss of generality we choose $\bar{b}=5$ as a specific example, and the corresponding range of $\bar{\alpha}$ is $-0.0642<\bar{\alpha}<0.025$. Gravitational KK resonances also exist in this kind of warp factor, but the situation will be different from case 1. The effective potential for different values of $\bar{\alpha}$ can be seen in Fig.~\ref{poalpha2}. From this figure we can see that there are three sub-wells, and the effect of $\bar{\alpha}$ to the effective potential is similar to that in case 1. We fix $\bar{z}_b=12$ and set $\bar{\alpha}$ as -0.0581, -0.0373, and 0.0161 as examples to study the gravitational KK resonances.
\begin{figure}
	\begin{center}
		\subfigure[The effective potential\label{potential2_1}]{\includegraphics[width=4.2cm]{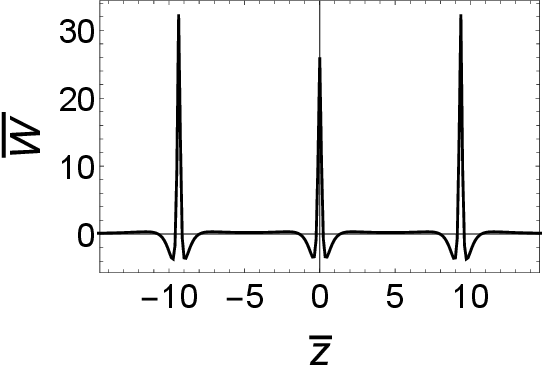}}
		\subfigure[The relative probability\label{probability2_1}]{\includegraphics[width=4.2cm]{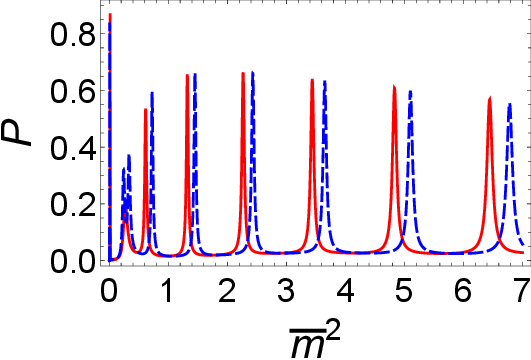}}
		\subfigure[The first odd wave function with $\bar{m}^2=0.0064$\label{resonant2_1odd}]{\includegraphics[width=4.2cm]{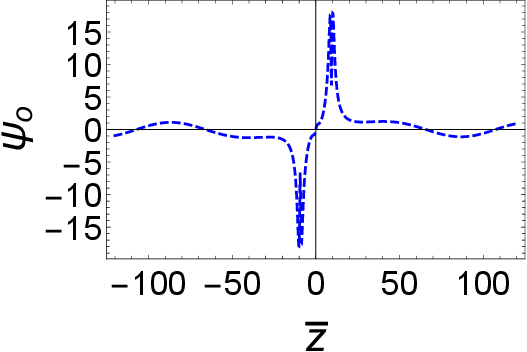}}
		\subfigure[The first even wave function with $\bar{m}^2=0.0137$\label{resonant2_1even}]{\includegraphics[width=4.2cm]{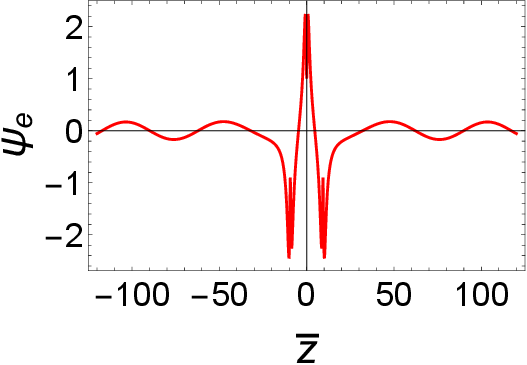}}
	\end{center}
	\caption{The effective potential, the relative probability, and wave functions of the first odd and even KK resonance with $\bar{b}=5$ and $\bar{\alpha}=-0.0581$ for case 2. \label{case2_1}}
\end{figure}
\begin{figure}
	\begin{center}
		\subfigure[The effective potential\label{potential2_2}]{\includegraphics[width=4.2cm]{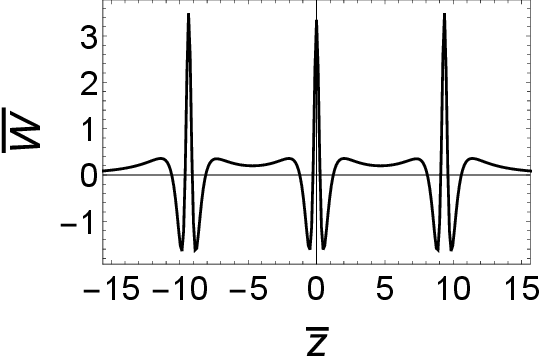}}
		\subfigure[The relative probability\label{probability2_2}]{\includegraphics[width=4.2cm]{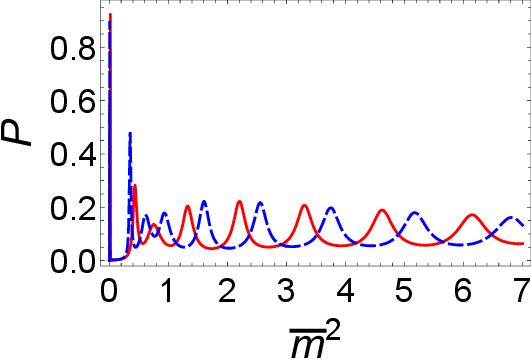}}
		\subfigure[The first odd wave function with $\bar{m}^2=0.0059$\label{resonant2_2odd}]{\includegraphics[width=4.2cm]{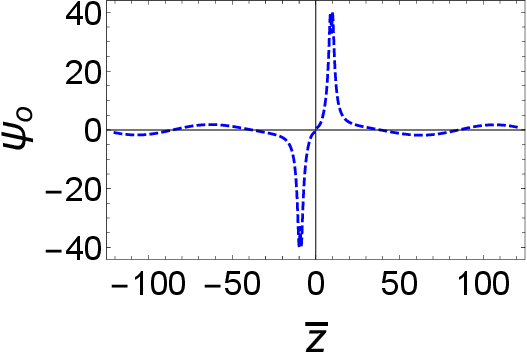}}
		\subfigure[The first even wave function with $\bar{m}^2=0.0172$\label{resonant2_2even}]{\includegraphics[width=4.2cm]{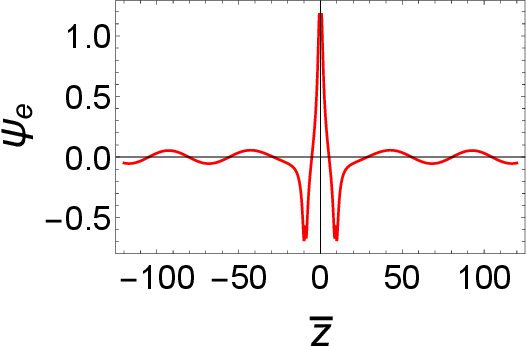}}
	\end{center}
	\caption{The effective potential, the relative probability, and wave functions of the first odd and even KK resonance with $\bar{b}=5$ and $\bar{\alpha}=-0.0373$ for case 2. \label{case2_2}}
\end{figure}
\begin{figure}
	\begin{center}
		\subfigure[The effective potential\label{potential2_3}]{\includegraphics[width=4.2cm]{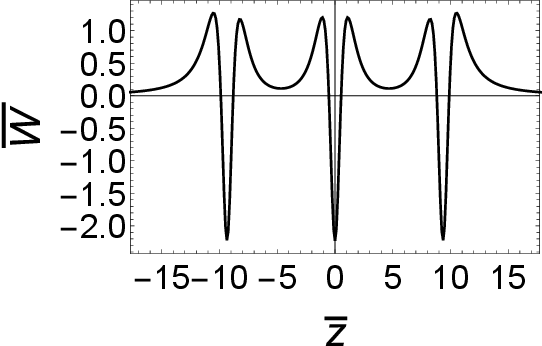}}
		\subfigure[The relative probability\label{probability2_3}]{\includegraphics[width=4.2cm]{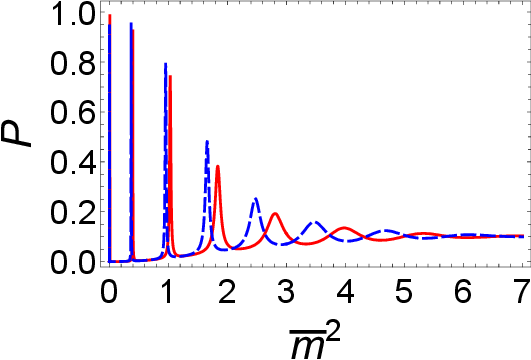}}
		\subfigure[The first odd wave function with $\bar{m}^2=0.0028$\label{resonant2_3odd}]{\includegraphics[width=4.2cm]{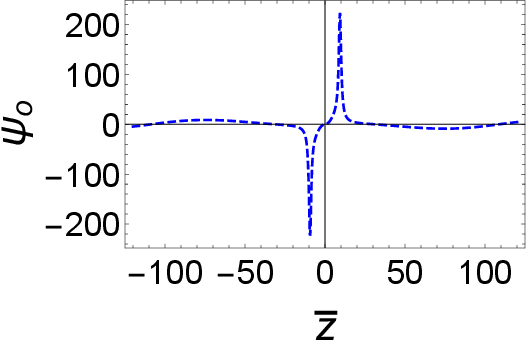}}
		\subfigure[The first even wave function with $\bar{m}^2=0.00892$\label{resonant2_3even}]{\includegraphics[width=4.2cm]{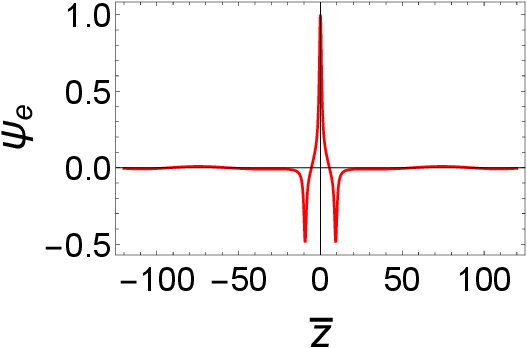}}
	\end{center}
	\caption{The effective potential, the relative probability, and wave functions of the first odd and even KK resonance with $\bar{b}=5$ and $\bar{\alpha}=0.016$ for case 2. \label{case2_3}}
\end{figure}

The shapes of the effective potential, the relative probability and the wave functions of the first odd and even KK resonances are shown in Figs.~\ref{case2_1}, \ref{case2_2}, and \ref{case2_3}. We can see that for some values of $\bar{\alpha}$, the relative probabilities of the KK resonances do not decrease with their masses $\bar{m}^2$ monotonously, which is similar to case 1. Besides, Figs.~\ref{case2_1} and \ref{case2_3} show that the masses of some KK resonances are very close, which is similar to the result in \cite{Xie2020}, where the doubly degenerate phenomenon happens. Comparing the effective potential with that in Ref. \cite{Xie2020}, we can see that they both have sub-structure, and this sub-structure is the main reason of this phenomenon.

\begin{figure}
	\begin{center}
		\subfigure[$\bar{m}^2$\label{malpha2}]{\includegraphics[width=4.2cm]{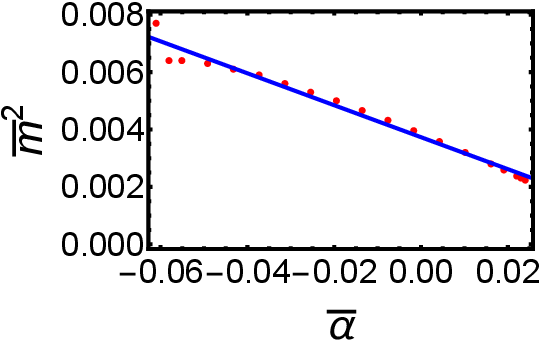}}
		\subfigure[$\bar{\alpha}$\label{taualpha2}]{\includegraphics[width=4.2cm]{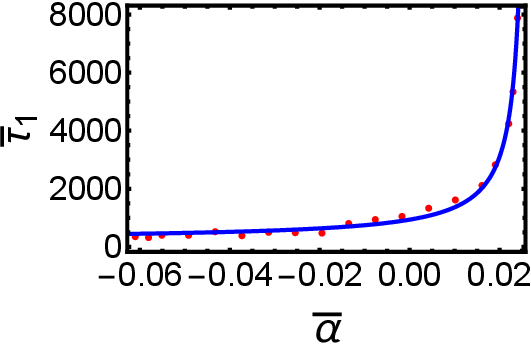}}
	\end{center}
	\caption{The effects of $\bar{\alpha}$ to the scaled mass $\bar{m}^2$ and the scaled lifetime $\bar{\tau}_1$ of the first KK resonance for case 2. Red dots are numerical results and solid blue lines are fit functions. In the left panel the fit function is $\bar{m}^2=0.00373-0.0557\bar{\alpha}$, in the right panel the fit function is $\bar{\tau}_1=\frac{1}{14.3-540\bar{\alpha}}+0.0238$.\label{fitfunction2}}
\end{figure}

We also investigate the lifetime $\bar{\tau}_1$ and the mass square $\bar{m}^2$ of the first resonance for different values of $\bar{\alpha}$. The results are shown in Fig.~\ref{fitfunction2} and we also get the following fit functions
\begin{align}
	\bar{m}^2&=0.00373-0.0557\bar{\alpha},\label{fitfunction2_1}\\
	\bar{\tau}_1&=\frac{10000}{14.3-540\bar{\alpha}}+238.\label{fitfunction2_2}
\end{align}
From Eq. \eqref{fitfunction2_1}, one can see that the mass square $\bar{m}^2$ of
the first KK resonance linearly decreases with the parameter $\bar{\alpha}$,
which is different from case 1. Equation~\eqref{fitfunction2_2} shows that
the lifetime increases with $\bar{\alpha}$ slowly first and then rapidly, just like the case 1.

From case 1 and case 2, we find that, when $\bar{\alpha}$ is small, there is an unusual phenomenon, i.e., the peak values of the relative probability of the KK resonances do not decrease monotonously with $\bar{m}^2$. This phenomenon appears when $\bar{\alpha} \lesssim -0.024$ for case 1 and $\bar{\alpha} \lesssim -0.014$ for case 2.

\subsection{Case 3}

\begin{figure}
	\begin{center}
		\subfigure[The warp factor\label{warp}]{\includegraphics[width=4.2cm]{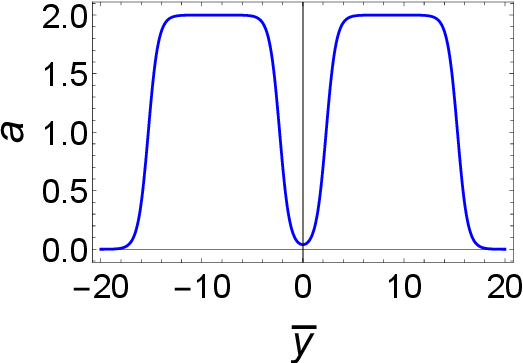}}
		\subfigure[The effective potential\label{drawing}]{\includegraphics[width=4.2cm]{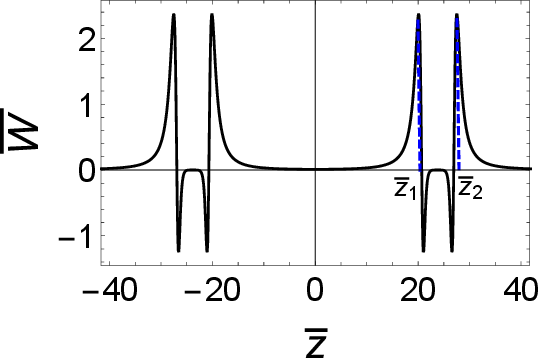}}
	\end{center}
	\caption{The warp factor and the effective potential $W(\bar{z})$ for case 3. The parameters are set as $\bar{b}=13$, $\bar{d}=2.3$, and $\bar{\alpha}=-0.00126$.\label{WarpFactorPotentialCase3}}
\end{figure}

In this case, we will focus on the gravitational KK resonances quasi-localized in sub-wells and between sub-wells. The warp factor is
\begin{align}
	a(\bar{y})=&\text{tanh}(\bar{y}-\bar{b}-\bar{d})-\text{tanh}(\bar{y}+\bar{b}+\bar{d})\nonumber \\
	&-\text{tanh}(\bar{y}+\bar{d})+\text{tanh}(\bar{y}-\bar{d}).
\end{align}
Compared with case 1, this warp factor has two platforms and the effective potential has more abundant sub-structure,  which can be seen in Fig.~\ref{WarpFactorPotentialCase3}.

From case 1 we have found that for large $\bar{b}$ the range of $\bar{\alpha}$ is $-0.219<\bar{\alpha}<0.00625$. In the current  case we set $\bar{b}=13$ and $\bar{d}=2.3$, for which the range of $\bar{\alpha}$ is still $-0.219<\bar{\alpha}<0.00625$.

\begin{figure}
	\begin{center}
		\includegraphics[width=6.2cm]{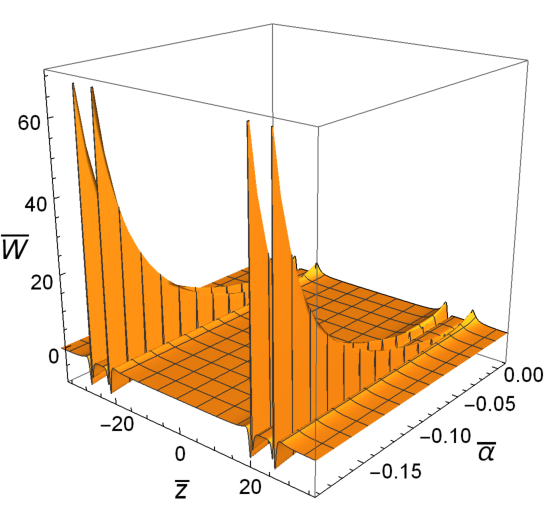}
	\end{center}
	\caption{The effect of the parameter $\bar{\alpha}$ on the effective potential with $\bar{b}=13$ and $\bar{d}=2.3$  for case 3.\label{potentialofalpha3}}
\end{figure}

The effective potential for different values of $\bar{\alpha}$ is shown in Fig.~\ref{potentialofalpha3}. In order to investigate the KK resonances only quasi-localized in sub-wells, we need to redefine the corresponding relative probability $P_s$ \cite{ZhongYi2019}:
\begin{equation}
P_{s}=\left\{\begin{array}{cl}
\frac{\int_{\bar{z}_{1}}^{\bar{z}_{2}}|\psi(\bar{z})|^{2} d \bar{z}}{\int_{\bar{z}_{m}-5\left(\bar{z}_{2}-\bar{z}_{1}\right)}^{\bar{z}_{m}+5\left(\bar{z}_{2}-\bar{z}_{1}\right)}|\psi(\bar{z})|^{2} d \bar{z}}, & \bar{z}_{m} \geq 5\left(\bar{z}_{2}-\bar{z}_{1}\right) \\
& \\
\frac{\int_{\bar{z}_{1}}^{\bar{z}_{2}}|\psi(\bar{z})|^{2} d \bar{z}}{\int_{0}^{10\left(\bar{z}_{2}-\bar{z}_{1}\right)}|\psi(\bar{z})|^{2} d \bar{z}}, & \bar{z}_{m}<5\left(\bar{z}_{2}-\bar{z}_{1}\right)
\end{array}\right. \label{subprobability}
\end{equation}
where $\bar{z}_1$ and $\bar{z}_2$ are the left and right edges of the right sub-well respectively (as Fig.~\ref{drawing} shows), and $\bar{z}_m=\frac{\bar{z}_1 + \bar{z}_2}{2}$. Besides, between two sub-wells there is also a potential well, and KK resonances can also be quasi-localized in it. We also need to redefine the relative probability in this region \cite{ZhongYi2019}:
\begin{equation}
P_{m}=\frac{\int^{\bar{z}_1}_{-\bar{z}_1}|\psi(\bar{z})|^2d\bar{z}}{\int^{10\bar{z}_1}_{-10\bar{z}_1}|\psi(\bar{z})|^2d\bar{z}}.\label{pmprobability}
\end{equation}


\begin{figure}
	\begin{center}
		\subfigure[The effective potential]{\includegraphics[width=4.2cm]{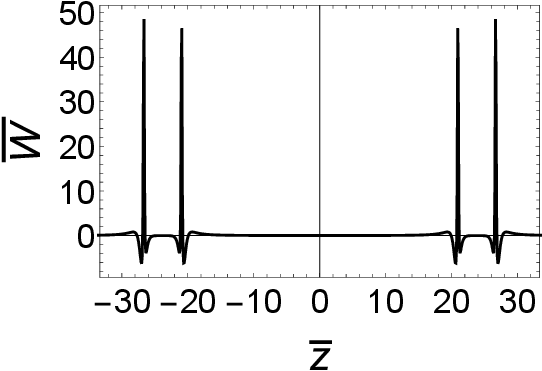}}
		\subfigure[The relative probability $P_m$ \label{Pbetween1}]{\includegraphics[width=4.2cm]{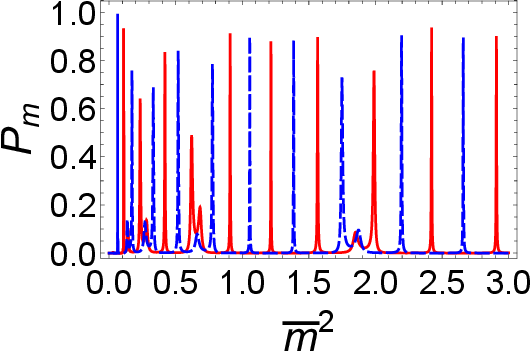}}
		\subfigure[The relative probability $P_s$ \label{Psub1}]{\includegraphics[width=4.2cm]{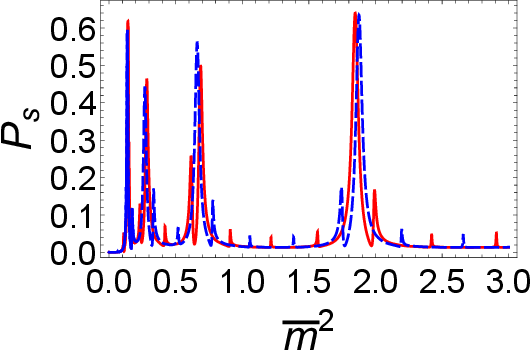}}
	\end{center}
	\caption{The effective potential, the relative probabilities of the even parity KK mode $\psi_e$ (dashed blue lines) and the odd parity KK mode $\psi_o$ (solid red lines) quasi-localized between the sub-wells $P_m$ and in the sub-wells $P_s$ with $\bar{b}=13$, $\bar{d}=2.3$, and $\bar{\alpha}=-0.189$ for case 3.\label{case3_1}}
\end{figure}

We solve the Shr\"odinger-like equation \eqref{Schrodingerlike} numerically with the boundary conditions \eqref{evencondition} and \eqref{oddcondition}. Using the redefined relative probabilities \eqref{subprobability} and \eqref{pmprobability}, one can get the gravitational KK resonances. Figures~\ref{case3_1}, \ref{case3_2}, and \ref{case3_3} show the effective potential and the corresponding relative probability for $\bar{\alpha}=-0.189$, $\bar{\alpha}=-0.0838$, and $\bar{\alpha}=-0.00126$, respectively. Through these figures we can see that the effects of $\bar{\alpha}$ on the effective potentials and relative probability are very considerable. Compared with the results of mimetic gravity brane world (see Fig.~11(c) in Ref.~\cite{ZhongYi2019}), we find that relative probabilities of KK resonances in sub-wells for small $\bar{\alpha}$ are much larger, which can be seen in Fig.~\ref{Psub1}. As $\bar{\alpha}$ approaches to zero, mimetic $f(R)$ gravity goes go back to mimetic gravity and the results of gravitational KK resonances also recover to that of in mimetic gravity \cite{ZhongYi2019}.

\begin{figure}
	\begin{center}
		\subfigure[The first odd wave function quasi-localized between sub-wells with $\bar{m}^2=0.0652$]{\includegraphics[width=4.2cm]{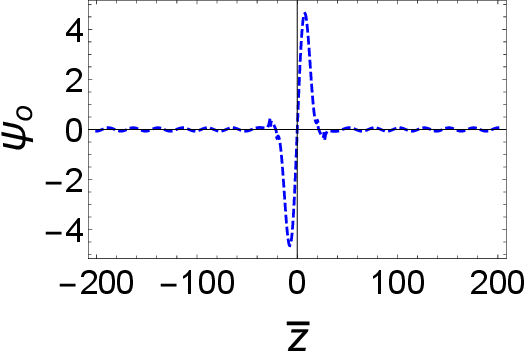}}
		\subfigure[The first even wave function quasi-localized between sub-wells with $\bar{m}^2=0.110$]{\includegraphics[width=4.2cm]{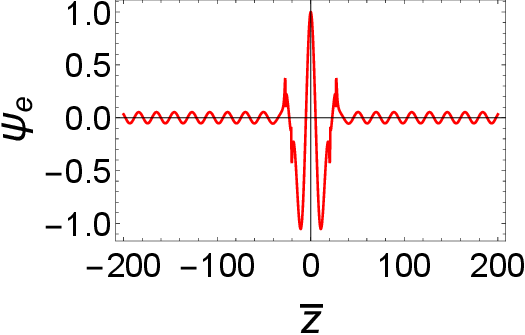}}
		\subfigure[The first odd wave function quasi-localized in sub-well with $\bar{m}^2=0.137$]{\includegraphics[width=4.2cm]{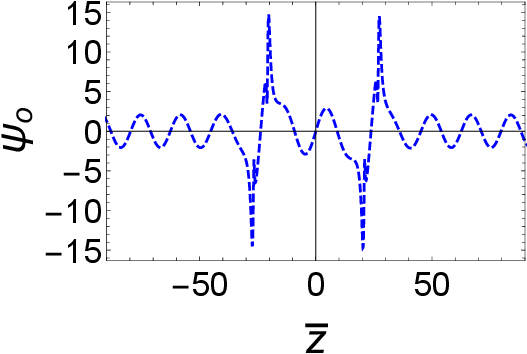}}
		\subfigure[The first even wave function quasi-localized in sub-well with $\bar{m}^2=0.142$]{\includegraphics[width=4.2cm]{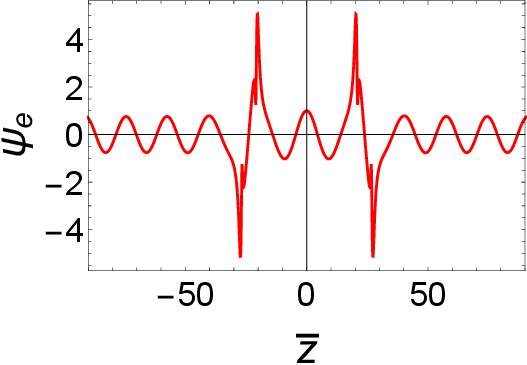}}
	\end{center}
	\caption{Resonances quasi-localized between sub-wells and in sub-wells with $\bar{b}=13$, $\bar{d}=2.3$, and $\bar{\alpha}=-0.189$ for case 3.\label{resonancesub1}}
\end{figure}

\begin{figure}
	\begin{center}
		\subfigure[The effective potential]{\includegraphics[width=4.2cm]{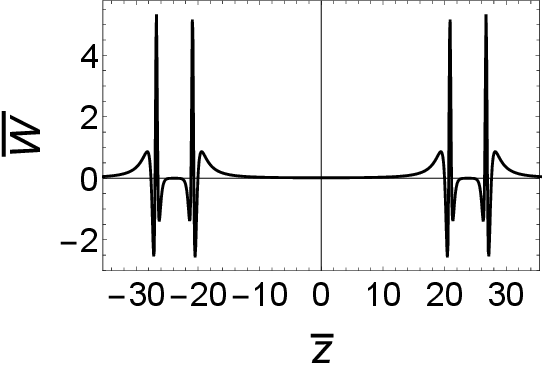}}
		\subfigure[The relative probability $P_m$ \label{Pbetween2}]{\includegraphics[width=4.2cm]{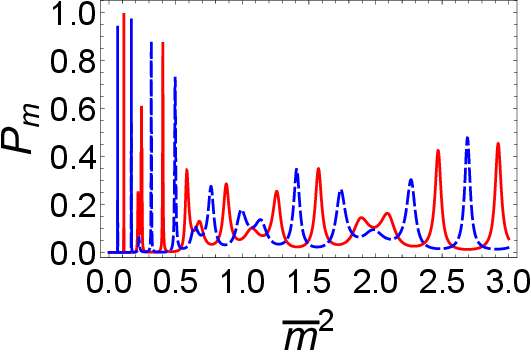}}
		\subfigure[The relative probability $P_s$ \label{Psub2}]{\includegraphics[width=4.2cm]{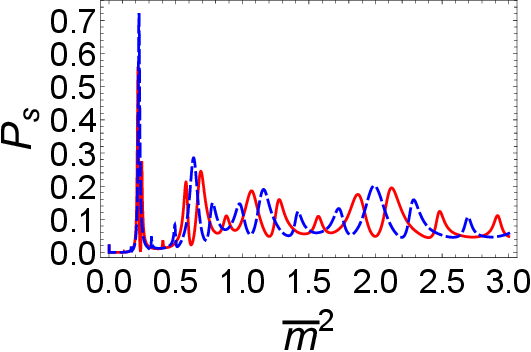}}
	\end{center}
	\caption{The effective potential, the relative probabilities of the even parity KK mode $\psi_e$ (dashed blue lines) and the odd parity KK mode $\psi_o$ (solid red lines) quasi-localized between the sub-wells $P_m$ and in the sub-wells $P_s$ with $\bar{b}=13$, $\bar{d}=2.3$, and $\bar{\alpha}=-0.0838$ for case 3.\label{case3_2}}
\end{figure}

\begin{figure}
	\begin{center}
		\subfigure[The first odd wave function quasi-localized between sub-wells with $\bar{m}^2=0.065$]{\includegraphics[width=4.2cm]{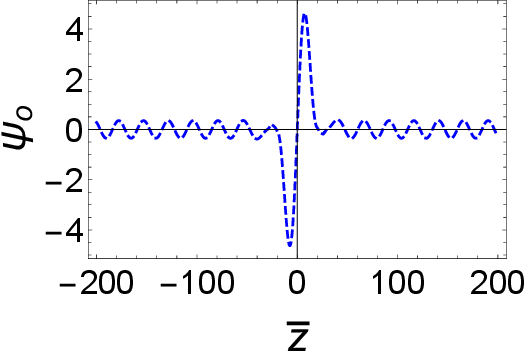}}
		\subfigure[The first even wave function quasi-localized between sub-wells with $\bar{m}^2=0.1112$]{\includegraphics[width=4.2cm]{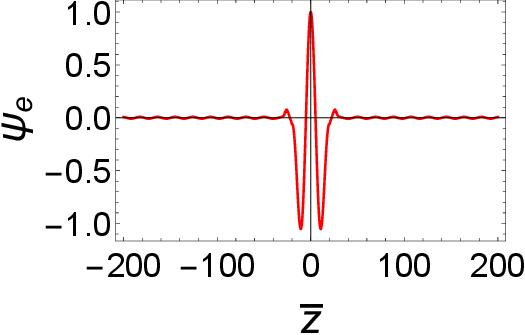}}
		\subfigure[The fist odd wave function quasi-localized in sub-well with $\bar{m}^2=0.225$]{\includegraphics[width=4.2cm]{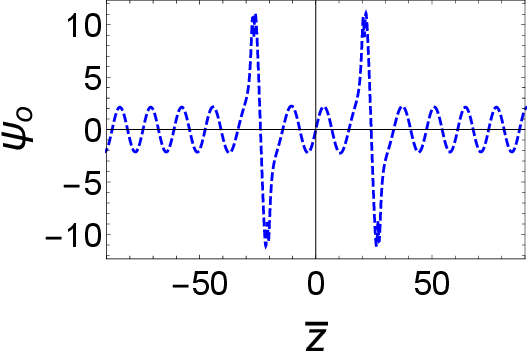}}
		\subfigure[The first even wave function quasi-localized in sub-well with $\bar{m}^2=0.217$]{\includegraphics[width=4.2cm]{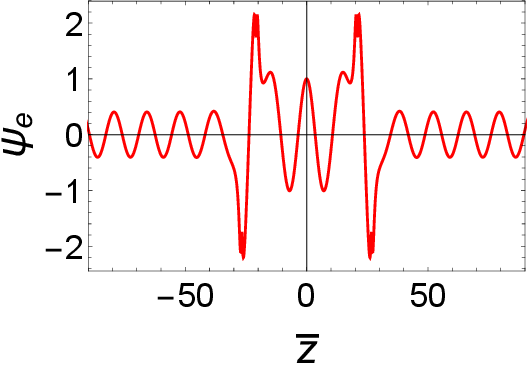}}
	\end{center}
	\caption{The KK resonances between sub-wells and in a sub-well with $\bar{b}=13$, $\bar{d}=2.3$, and $\bar{\alpha}=-0.0838$ for case 3.\label{resonancesub2}}
\end{figure}

\begin{figure}
	\begin{center}
		\subfigure[The effective potential]{\includegraphics[width=4.2cm]{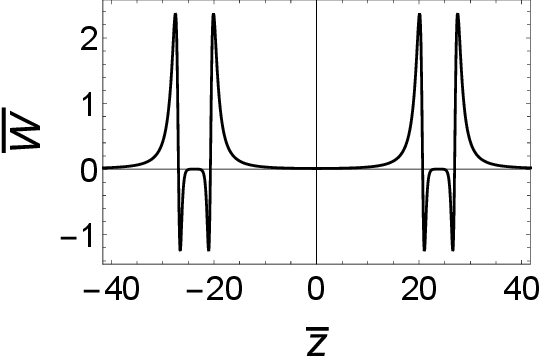}}
		\subfigure[The relative probability $P_m$ \label{Pbetween3}]{\includegraphics[width=4.2cm]{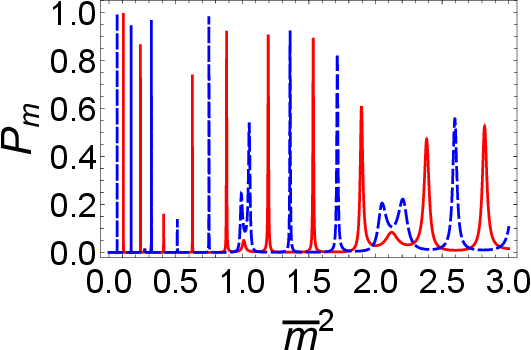}}
		\subfigure[The relative probability $P_s$ \label{Psub3}]{\includegraphics[width=4.2cm]{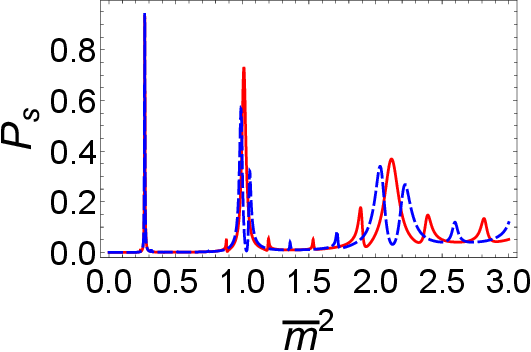}}
	\end{center}
	\caption{The effective potential, the relative probabilities of the even parity KK mode $\psi_e$ (dashed blue lines) and the odd parity KK mode $\psi_o$ (solid red lines) quasi-localized between the sub-wells $P_m$ and in the sub-wells $P_s$ with $\bar{b}=13$, $\bar{d}=2.3$, and $\bar{\alpha}=-0.00126$ for case 3.\label{case3_3}}
\end{figure}

\begin{figure*}
	\begin{center}
		\subfigure[The first odd wave function quasi-localized between sub-wells with $\bar{m}^2=0.0608$]{\includegraphics[width=4.2cm]{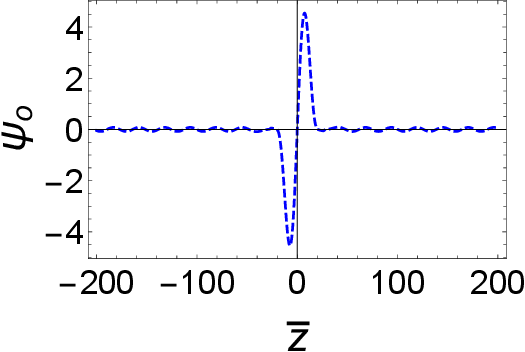}}
		\subfigure[The first even wave function quasi-localized between sub-wells with $\bar{m}^2=0.108$]{\includegraphics[width=4.2cm]{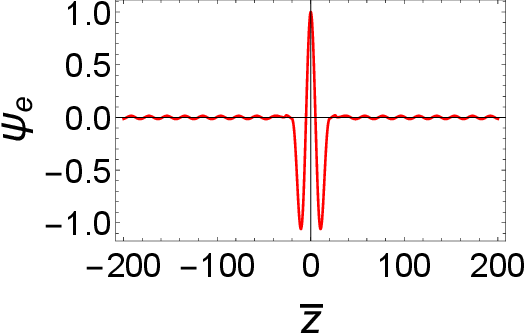}}
		\subfigure[The fist odd wave function quasi-localized in sub-well with $\bar{m}^2=0.268$]{\includegraphics[width=4.2cm]{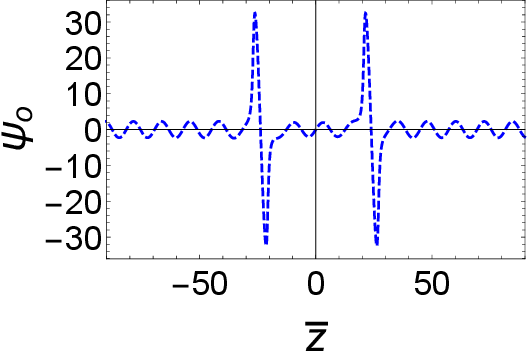}}
		\subfigure[The first even wave function quasi-localized in sub-well with $\bar{m}^2=0.268$]{\includegraphics[width=4.2cm]{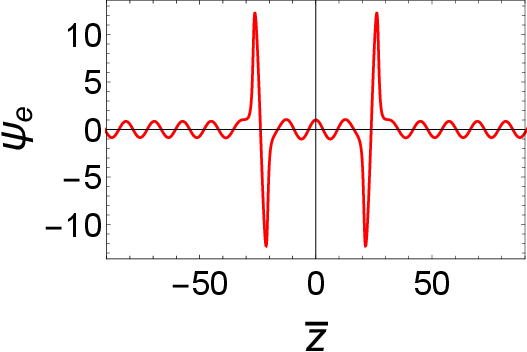}}
	\end{center}
	\caption{The KK resonances between sub-wells and in a sub-well with $\bar{b}=13$, $\bar{d}=2.3$, and $\bar{\alpha}=-0.00126$ for case 3.\label{resonancesub3}}
\end{figure*}

From Figs.~\ref{Psub1}, \ref{Psub2}, and \ref{Psub3} we can see that the peaks of relative probabilities of gravitational KK resonances in the sub-wells appear in clusters. And the main peak value of each cluster does not monotonically decrease with $\bar{m}^2$ when $\bar{\alpha}$ is small, this is because the sub-wells also have their sub-structure changing with $\bar{\alpha}$. From Figs.~\ref{Pbetween1}, \ref{Pbetween2}, and \ref{Pbetween3} we can also see that the peak values of relative probabilities of gravitational KK resonances quasi-localized between sub-wells will not monotonically decrease with $\bar{m}^2$. Besides, comparing the above two cases, we see that there are no corresponding KK resonances with large relative probabilities $P_m$ at the locations of the main peaks of clusters in $P_s$. This is caused by the existence of sub-structure of the effective potential. The wave functions of KK resonances quasi-localized both between sub-wells and in sub-wells are shown in Figs.~\ref{resonancesub1}, \ref{resonancesub2}, and \ref{resonancesub3} for corresponding $\bar{\alpha}$.

On the other hand, from Figs.~\ref{Psub1}, \ref{Psub2}, and \ref{Psub3} we find that the mass square $\bar{m}^2$ of the first odd parity KK resonance and the first even parity KK resonance are very close. One could also regard this as doubly degenerate \cite{Xie2020}.

\section{Conclusions and discussions}\label{sec_5}

In this work, we studied the brane world system in mimetic $f(R)$ gravity. Due to the existence of the mimetic scalar field, we can get thick branes with abundant inner structure. We gave three kinds of solutions of the thick brane model for $f(R)=R+\alpha R^2$.

Then we investigated the stability of mimetic $f(R)$ brane world, and found that the system is stable under the tenser perturbation. The existence of the normalized graviton zero mode indicates that four-dimensional gravity can be restored. The result is same as the tensor perturbation of $f(R)$-branes in Ref.~\cite{ZhongYuan2011}, this is because that the tenser perturbation is independent of the mimetic scalar field.

After that, we studied the gravitational KK resonances, which are quasi-localized on the brane. We only studied the effects of the parameter $\alpha$ on the behavior of the KK resonances. We found that for a small $\alpha$, the effective potential has a sub-structure which leads to an unusual phenomenon in the KK resonance spectrum. The lifetime of the first KK resonance increases rapidly  while $\alpha$ close to its upper bound. In case 3 we focused on gravitational KK resonances quasi-localized in sub-wells and between sub-wells. We found that KK  resonances can be quasi-localized both in sub-wells and between sub-wells. If $\alpha$ tends to zero the result would tend to mimetic gravity brane world in Ref.~\cite{ZhongYi2019}. Compared with Ref.~\cite{ZhongYi2019}, the relative probabilities of KK resonances in the sub-wells are more obvious and the main peak values of the clusters do not monotonically decrease with $\bar{m}^2$ for a small $\alpha$. Besides, the phenomenon of gravitational KK resonance doubly degenerate also appears when we studied KK resonances quasi-localized in a sub-well.

We note that it is proper to set the five-dimensional fundamental mass scale $M_*$ as unity rather than the Planck mass. In this model the relation between the four-dimensional spacetime mass scale $M_{\text{Pl}}$ and five-dimensional spacetime fundamental mass scale $M_*$ is as follows
\begin{equation}
	M^2_{\text{Pl}}=M_*^3 \int^{\infty}_{-\infty}a(z)^3 f_R dz,
\end{equation}
which means the four-dimensional mass scale is effective. The five-dimensional fundamental scale is determined by the warp factor and $f(R)$. And we use the parameter $k$ to recover dimensions of parameters $y$, $z$, $b$, $d$, $\tau$, $\Gamma$, $m$, $W$, and $\alpha$. For example, if we set $k=M_*$, for model 2, we have
\begin{align}
	M_{\text{Pl}}^2=&M_*^3 \Bigg( \frac{2}{3 k} \text{csch}^2(2 b k) \Big(-\left(16 \alpha  k^2+3\right) \cosh (4 b k) \nonumber
	\\
	&+6 b k \big(16 \alpha  k^2 \tanh (b k) \left(\coth ^2(b k)+1\right)+\sinh (4 b k)\big)\nonumber \\
	&-80 \alpha  k^2+3 \Big) \Bigg).  \label{fundamental}
\end{align}
If we set the parameters $kb=10$ and $k^2 \alpha=-0.008$ then the five-dimensional spacetime fundamental mass scale can be obtain as $M_*=M_{\text{Pl}}/8.73=1.40 \times 10^{15}~\text{TeV}$ from Eq.~\eqref{fundamental}. And the parameter $b= 1.41 \times 10^{-33}~ \text{m}$. When the energy of a particle approaches to the fundamental mass scale the effect of quantum gravity will become obvious. In this model the effect of quantum gravity is not considered. From Fig.~\ref{case1_1} the mass of the first massive graviton is $m=\sqrt{0.0831} \times 1.40 \times 10^{15}~\text{TeV}=1.16 \times 10^{14}~\text{TeV}$, which is less than the fundamental mass scale. So the massive graviton can product in a collider even its energy lower than the fundamental mass scale.

At last, we discuss briefly the correction of the four-dimensional Newtonian potential from the gravitational KK modes. For the brane located around $z=z_0$, the four-dimensional Newtonian potential correction of a mass point $M$ is given by \cite{Csaki2000}
\begin{equation}
	U_{N}(r) \sim G_{N} \frac{M}{r}\left[1+\int_{0}^{\infty} d m~ \mathrm{e}^{-m r} \psi_{m}^{2}\left(z_{0}\right)\right],
\end{equation}
where $r$ is the distance to the mass point on the brane, and $\psi_m(z)$ is the normalized wave function with mass $m$ and the normalized constant $N_m$. In this paper if the brane is located around $z=0$, the contribution to the four-dimensional Newtonian potential is only from even KK modes. For the brane with inner structure the KK resonances that can be quasi-located at sub-brane also have contribution to the four-dimensional Newtonian potential. From Ref.~\cite{Csaki2000} we know that the normalization constant $N_m$ is decided by the asymptotic behavior of the effective potential $W(z)$ at the boundary of the extra dimension. For the solutions considered in this paper, if we assume that $W(z)\propto \frac{\beta(\beta+1)}{z^2}$ and $\psi_m(0)\sim (m/k)^{\beta-1}$ at $z\gg 1/k$, we find $\beta=\frac{3}{2}$, which is not associated with $\alpha$. Hence, for the brane located around $z=0$ the correction to the four-dimensional Newtonian potential is
\begin{equation}
	U_{N}(r) \sim G_{N} \frac{M}{r}\left[1+\frac{C}{(kr)^2}\right],
\end{equation}
where $C$ is a dimensionless constant determined by the structure of brane. Here, we only consider the asymptotic behavior of the effective potential $W(z)$ at the boundary of the extra dimension. The inner structure of the effective potential around $z=0$ should have an effect on this result. This is a difficult issue and it will be studied in our future work.

In the model of mimetic brane world with abundant inner structure, it is also interesting to investigate localization of various kinds of matter fields and the corresponding KK resonance structures, which will also be studied in our future work.


\section{Acknowledgments}

We would like thank Yi Zhong, Zheng-Quan Cui, Hao Yu, Bao-Min Gu, and Qin Tan for helpful discussions. This work was supported in part by the National Key Research and Development Program of China (Grant No. 2020YFC2201503), the National Natural Science Foundation of China (Grants No. 11875151, No. 11522541, and No. 12047501), the 111 Project (Grant No. B20063), and the Fundamental Research Funds for the Central Universities (Grants No. lzujbky-2019-ct06). W. D. Guo was supported by the scholarship granted by the Chinese Scholarship Council (CSC).


\end{document}